\documentclass[twocolumn,mathlines]{aastex631}

\usepackage{newtxtext,newtxmath}
\usepackage[T1]{fontenc}
\usepackage{graphicx}	
\usepackage{amsmath}	

\newcommand{\<}{\langle}
\renewcommand{\>}{\rangle}
\newcommand{\eee}{\equiv}

\newcommand{\eref}[1]{Eq.~(\ref{#1})}

\newcommand{\Lenstool}{\text{\sc{Lenstool}}}

\newcommand{\nug}{\nu_\text{gal}}
\newcommand{\Ngal}{N_\text{gal}}

\newcommand{\sigmalos}{\sigma_\text{los}}
\newcommand{\sigmar}{\sigma_\text{r}}

\newcommand{\Sigmag}{\Sigma_\text{gal}}

\newcommand{\vrf}{v_\text{rf}}
\newcommand{\vesp}{v_\text{esp}}

\newcommand{\zspec}{z_\text{spec}}

\newcommand{\Ls}{L_*}

\newcommand{\rts}{r_{\text{t}*}}

\newcommand{\rs}{r_\text{s}}

\newcommand {\sref}[1]        {Section~\ref{#1}}
\newcommand {\aref}[1]        {Appendix~\ref{#1}}
\newcommand {\fref}[1]        {Fig.~\ref{#1}}
\newcommand {\tref}[1]        {Table~\ref{#1}}

\begin{document}


\title{Constraining the Nature of Dark Matter from Tidal Radii of Cluster Galaxy Subhalos}

\author[0000-0002-9370-4490]{Barry T. Chiang}
\affiliation{Department of Astronomy, Yale University, New Haven, CT 06511, USA}

\email{barry.chiang@yale.edu}

\author[0000-0001-7040-4930]{Isaque Dutra}
\affiliation{Department of Physics, Yale University, New Haven, CT 06511, USA}

\author[0000-0002-5554-8896]{Priyamvada Natarajan}
\affiliation{Department of Astronomy, Yale University, New Haven, CT 06511, USA}
\affiliation{Department of Physics, Yale University, New Haven, CT 06511, USA}
\affiliation{Yale Center for the Invisible Universe, Yale University, New Haven CT 06511, USA}


\begin{abstract}
Gravitational lensing by galaxy clusters provides a powerful probe of the spatial distribution of dark matter and its microphysical properties. Strong and weak lensing constraints on the density profiles of subhalos and their truncation radii offer key diagnostics for distinguishing between collisionless cold dark matter (CDM) and self-interacting dark matter (SIDM). Notably, in the strongly collisional SIDM regime, subhalo core collapse and enhanced mass loss from ram-pressure stripping predict steeper central density slopes and more compact truncation radii\textemdash features that are directly testable with current lensing data. We analyze subhalo truncation in eight lensing clusters (Abell 2218, 383, 963, 209, 2390, and MACS J0416.1, J1206.2, J1149.6) that span the redshift range $\<\zspec\> \simeq 0.17$\textendash$0.54$ with virial masses $M_{200} \simeq0.41$\textendash$2.2\times 10^{15}$~M$_\odot$ to constrain SIDM versus CDM. Our results indicate that the outer spatial extents of subhalos are statistically consistent with CDM, corroborated by redshift- and mass-matched analogs from the Illustris-TNG simulations. We conclude that the tidal radii of cluster galaxy subhalos serve as an important and complementary diagnostic of the nature of dark matter in these violent, dense environments.  
\end{abstract}



\section{Introduction}\label{sec:Introduction}
    Numerous independent astrophysical probes have been used to constrain the nature of dark matter. The standard cold dark matter paradigm (CDM) has been remarkably successful in accounting for observations on a wide range of scales \citep{Bahcall2015}. However, over the past two decades, several major tensions, namely inconsistencies with the CDM model, have been reported on small scales within galaxies, including the core-cusp problem, missing satellite problem, too-big-to-fail problem, plane-of-satellites problem, and the diversity of rotation curves \citep{Weinberg2015PNAS11212249W, Bullock2017ARAA55343B, DelPopolo2017Galax517D, Sales2022NatAs6897S}. Various solutions, with the improved treatments of galaxy formation and assembly in simulations, have by and large addressed and mitigated most of these small-scale issues. For example, the implementation of baryonic feedback processes \citep{ElZant2001ApJ560636E, Goerdt2010ApJ7251707G}; dynamical friction \citep{Pontzen2012MNRAS4213464P, Teyssier2013MNRAS4293068T, Freundlich2020MNRAS4914523F}; and/or subhalo velocity anisotropy \citep{Chiang2024arXiv241103192C} have all been invoked to naturally explain the presence of cored central density profiles in observed dwarf galaxies.
    
    Similarly, viable solutions to alleviate and address the other small-scale ``controversies'' that CDM predicts \citep[e.g.][]{Homma2024PASJ76733H, Ostriker2019ApJ88597O, Sawala2023NatAs7481S} remain generally compatible with observations given the current uncertainties in understanding galaxy formation models and the numerical resolution limit of cosmological simulations \citep[e.g.][]{Brooks2013ApJ76522B, Tomozeiu2016ApJ827L15T, Verbeke2017AA607A13V, Lovell2017MNRAS4682836L, Sawala2016MNRAS4571931S}.

    On galaxy cluster scales, (sub)halos in these denser cosmic environments produce measurable individual gravitational lensing effects that are now detectable in deeper Hubble and James Webb Space Telescope images, offering yet another independent and stringent set of consistency tests of dark matter models. Observed strong and weak gravitational lensing effects have permitted high resolution mapping of the dark matter distribution in these massive clusters. This has enabled a detailed comparison of the lensing derived properties of cluster dark matter subhalos with the concrete predictions from CDM simulations. Two new notable tensions with CDM have recently been reported from this exercise. The first arises when comparing the radial distribution of subhalos from these cluster lensing mass models with CDM simulations \citep{Natarajan2017MNRAS4681962N} and the second arises when quantifying the probability of strong lensing events produced by the dark matter substructure, namely the Galaxy–Galaxy Strong Lensing (GGSL) probability. A deficit in the number of subhalos was found in the inner regions compared to observational data in the redshift and mass-matched cluster analogs of the IllustrisTNG simulation suite \citep{Natarajan2017MNRAS4681962N}. And \citet{Meneghetti:2020yif} found that the GGSL probability predicted by CDM cosmological simulations falls more than an order of magnitude below the observationally determined value from a sample of 11 well-studied cluster lenses that reveal background lensed sources to $z \simeq 7$. The GGSL discrepancy suggests that the inner regions of the subhalos are significantly more efficient lenses than predicted by CDM. At the moment, there appears to be no obvious resolution of these discrepancies within the CDM paradigm \citep{Meneghetti:2022apr, Ragagnin:2022usu,Dutra2025ApJ97838D}, even when baryonic effects such as extreme contraction and compactification are invoked to steepen the subhalo density profile produced by assuming even extremely unrealistically strong baryonic processes \citep{Tokayer:2024wwo}. However, a minimal extension of CDM to include generic dark matter self-interaction \citep[e.g.][]{Feng:2009mn, Hall:2009bx, Tulin:2012wi, Aboubrahim:2020lnr}, appears to largely alleviate this tension. In fact, \citet{Dutra2025ApJ97838D} show that rearranging the mass within the innermost regions of cluster subhalos can fully account for this mismatch and resolve this tension. In the strongly collisional regime of self-interacting dark matter (SIDM) models, subhalos undergoing core collapse can exhibit central density slopes (significantly) steeper than $\rho(r\rightarrow0)\propto r^{-1}$ of the typical NFW profile \citep{Navarro:1996gj}. As noted in \citet{Yang2021PhRvD104j3031Y}, this potential solution offers a way to increase the probability of strong lensing cross section, and \citep{Dutra2025ApJ97838D} explicitly demonstrated that with collapsed core subhalos with a central density slope of around $\rho(r\rightarrow0)\propto r^{-2.9}$ in the inner-most regions, the GGSL tension could be fully resolved.

    In this work, we explore an orthogonal diagnostic that includes the combination of the strong and weak lensing regimes to further stress test CDM, once again in galaxy clusters and from subhalos within them. Here we focus instead on the outer regions of subhalos and their properties, that might telegraph the nature of dark matter via the imprint of the tidal forces that act on them. For instance, subhalos comprised of dark-matter particles with non-trivial self-interaction, like core-collapse SIDM for instance, would incur additional ram-pressure stripping that, in the strongly collisional regime, will dominate over the purely gravitational tidal mass loss predicted for collisionless CDM subhalos \citep[e.g.][]{Moore:2000fp, Furlanetto:2001tw}. Therefore, the subhalo tidal truncation radii of cluster member galaxies would be more compact and, in this instance, would depend sensitively on the dark matter self-interaction cross-section. This potential phenomenological signature was first explored by \citet{Natarajan:2002cw} for the cluster lens Abell 2218, resulting in a conservative $5\sigma$ exclusion bound on the dark-matter self-interaction cross section $\sigma_\text{SIDM}/m_\text{SIDM} \lesssim 42 \text{cm}^2/\text{g}$ from the truncation radius distribution of 25 spectroscopically confirmed cluster member galaxies. Here we revisit this test of the nature of dark matter that is telegraphed in the sizes of the truncation radii of cluster subhalos using significantly more sophisticated and well constrained lensing mass models.

    Given the plethora of high-fidelity galaxy cluster lensing measurements from HST observational programs like CLASH \citep{Postman2012ApJS19925P, Umetsu2014ApJ795163U}, \textit{HST Frontier Fields} \citep{Lotz2017ApJ83797L, Natarajan2017MNRAS4681962N}, and RELICS \citep{Coe2019ApJ88485C, Cerny2018ApJ859159C} as well as the highly complete spectroscopic catalogs of associated cluster member galaxies derived from multi-object spectrographs like MUSE (e.g. see recent review \citet{Natarajan2024SSRv22019N} and references therein). Here, we expand upon this initial analysis to survey eight massive galaxy clusters spanning redshifts $z \simeq 0.17$\textendash$0.54$. We explore the statistical compatibility between empirically inferred and CDM/SIDM-predicted outer tidal extents of subhalos, leveraging these richer datasets currently available with a larger number of spectroscopically confirmed cluster-member galaxies and the resulting significantly higher quality cluster mass distributions reconstructed from combining strong and weak lensing data.
    
    This paper is organized as follows: \sref{sec:Cluster_DM} describes the Bayesian optimization process used to construct parametric cluster lensing mass models and the inference of subhalo tidal extents from them performed using \Lenstool~\citep{Natarajan1997MNRAS287833N,Kneib2011A&ARv1947K}. The properties of the lensing cluster sample and those of the associated cluster members studied here are summarized in \sref{sec:Cluster_Sample}. In \sref{sec:rt_Estimates}, we first describe and compute analytic estimates of the subhalo tidal truncation radii for both CDM and SIDM; and present the cross-validation of CDM estimates with cosmological simulations. The comparison between CDM and SIDM predictions against lensing inferred values of subhalo truncation radii are presented in \sref{sec:rt}. The uncertainties and limitations in lensing models, assumptions and adopted procedures are discussed in \sref{sec:rt_Uncertainties}. We discuss the implications of our findings in \sref{ssec:rt_Implications} and conclude in \sref{sec:Conclusions}.

    We clarify that in this work, $R$ denotes the 2D projected radius and $r$ the 3D (deprojected) radius from a center of reference. Aligned with most previous work on cluster lensing measurements, we adopt the following cosmological parameters when required: a flat $\Lambda$CDM cosmology with $H_0 = 70$~km~s$^{-1}$~Mpc$^{-1}$, $\Omega_\text{m} = 0.3$, and $\Omega_\Lambda = 0.7$, giving $t_\text{Universe} = 13.46$~Gyr.

\section{Mass distributions derived from cluster lensing}\label{sec:Cluster_DM}

In this section, we describe the highly flexible parametric mass models derived from combining strong- and weak-lensing observations of massive cluster lenses that are particularly well suited for direct comparison with cosmological simulation. This modeling methodology, by virtue of the adopted conceptual framework and self-similar parametric models deployed for the cluster members, naturally provides constraints on the tidal extents of the subhalos hosting cluster member galaxies as outlined in \sref{ssec:Cluster_DM_rt}. We specifically choose parametric models as partitioning the total mass distribution as a sum of larger scale halos and smaller scale subhalos as done in this work aligns well with the conceptual framework of halo and subhalo catalogs adopted in CDM simulations.

\subsection{Optimized lensing mass models}\label{ssec:Cluster_DM_Lenstool}

The commonly used, publicly available software package \Lenstool\footnote{\url{https://projets.lam.fr/projects/lenstool/wiki}}~offers an efficient way to use the observed lensing signals, namely combine the positions and brightnesses of the multiply-imaged strongly lensed galaxies, with the positions and shapes of the weakly lensed background galaxies to reconstruct the detailed mass distribution of massive lensing clusters. These standard methods have been in use for over two decades and have been tested against simulated clusters from multiple independent cosmological simulation suites. A review of these modeling methods, their power, and limitations can be found in \cite{Natarajan+2024, Meneghetti2017MNRAS4723177M, Kneib2011A&ARv1947K}. In essence, \Lenstool~performs the multi-scale Bayesian optimization to reconstruct cluster lens mass distributions as the linear superposition of parametric profiles that are constrained by strong and weak lensing data to yield mass maps \citep{2011ascl.soft02004K}:
\begin{align}
	\phi_\text{tot} = \sum_i \phi^\text{halo}_i + \sum_j \phi^\text{subhalo}_j +  \phi_\kappa, 
\end{align}
where $\phi^\text{halo}_i$ represents Mpc-scale halos associated with the smoother larger-scale cluster gravitational potential; $\phi^\text{subhalo}_j$ the kpc-scale subhalos associated with cluster member galaxies, and $\phi_\kappa$ a potential constant external shear field, according to the conceptual model in \citet{Natarajan1997MNRAS287833N}. Unless specified otherwise, the (sub)halos are modeled by self-similar dual pseudoisothermal elliptical mass distributions (dPIE) whose 3D density profile $\rho_\text{dPIE}$, enclosed mass profile $M_\text{dPIE}$, and 2D surface density profile $\Sigma_\text{dPIE}$ are 
\begin{align}\label{eqn:rho_dPIE}
		&\rho_\text{dPIE}(r) \eee \frac{\rho_0}{\Big(1+\frac{r^2}{r^2_\text{core}}\Big)\Big(1+\frac{r^2}{r^2_\text{t}}\Big)},\\
        &M_\text{dPIE}(r) = 4\pi\rho_0\bigg(\frac{r_\text{core}^2 r_\text{t}^2\big[r_\text{t} \tan^{-1}\big(\frac{r}{r_\text{t}}\big)-r_\text{core} \tan^{-1}\big(\frac{r}{r_\text{core}}\big)\big]}{r_\text{t}^2-r_\text{core}^2}\bigg),\nonumber \\
		&\Sigma_\text{dPIE}(R) = \frac{\Sigma_0 r_\text{core}}{1-\big(\frac{r_\text{core}}{r_\text{t}}\big)}\bigg(\frac{1}{\sqrt{\smash[b]{r_\text{core}^2}+R^2}}-\frac{1}{\sqrt{\smash[b]{r_\text{t}^2}+R^2}}\bigg),\nonumber
\end{align}
where $r_\text{core}$ denotes the core radius, and $r_\text{t}$ tidal truncation radius. One attractive feature of this choice of model parameterization is that the total mass $M_\text{dPIE}(r\rightarrow\infty) = 2\pi \Sigma_0 r_\text{core} r_\text{t}$ is finite.

For each subhalo, the normalization coefficients $\rho_0$ and $\Sigma_0$ are set uniquely by the effective velocity dispersion: 
\begin{align}\label{eqn:sigma_dPIE}
    \sigma_\text{dPIE} &\eee \frac{4 G\pi \rho_0}{3}\frac{r_\text{core}^2 r_\text{t}^3}{(r_\text{t}-r_\text{core})(r_\text{t}+r_\text{core})^2}
    = \frac{4G \Sigma_0}{3}\frac{r_\text{core} r_\text{t}^2}{r_\text{t}^2-r_\text{core}^2},
\end{align}
where $G$ denotes the gravitational constant. The numerical values of $\sigma_\text{dPIE}$ are derived for all substructures simultaneously by optimizing the entire observed cluster lens image. It turns out that $\sigma_\text{dPIE}$ is related to the physical central velocity dispersion of each member galaxy by $\sigma_\text{dPIE}^2 = \frac{2}{3}\sigma_\text{gal}^2$. Lastly, in projection, each substructure is allowed to have non-zero ellipticity (e.g. see Sec.~3.1 of \citet{Dutra2025ApJ97838D}).


\subsection{Empirically constrained subhalo tidal radii}\label{ssec:Cluster_DM_rt}

With the further assumption that light traces mass (on the mass and spatial scales of interest), the free parameters associated with individual subhalos are constrained by the empirical scaling relation of cluster member galaxies hosted in them \citep{Natarajan:2002cw, Eliasdottir2007arXiv07105636E,Limousin2024}
\begin{align}\label{eqn:E_Scaling_Relations}
	\begin{aligned}
        &\sigma_\text{dPIE} = \sigma_{\text{dPIE}*}\bigg(\frac{L}{\Ls}\bigg)^\alpha,\\
		&r_\text{t} = \rts\bigg(\frac{L}{\Ls}\bigg)^\beta,\\
	&r_\text{core} = r_{\text{core}*}\bigg(\frac{L}{\Ls}\bigg)^{1/2},
    \end{aligned}
\end{align}
where quantities with a star symbol subscript denote the characteristic member galaxy properties, obtained by fitting a Schechter function to the luminosties of the hosted member galaxies \citep{Schechter1976ApJ203297S}. The values $\alpha = 0.25$ and $\beta=0.5$ corresponding to the Faber\textendash Jackson relation \citep{Faber1976ApJ204668F} are favored in all the best-fit cluster mass models of our sample \tref{tab:Cluster_Properties}, bar one where we find a best-fit $\alpha = 0.28$ and $\beta = 0.64$ (see model details below).

From \eref{eqn:sigma_dPIE}, the total mass of individual substructure scales as $M_\text{dPIE}(r\rightarrow\infty) = (9\sigma_{\text{dPIE}*}^2 r_{\text{t}*}/2G)(L/L_*)^{2\alpha+\beta}$, yielding a mass-to-light ratio that scales as $\Upsilon \propto(L/L_*)^{2\alpha+\beta-1}$. Physically, $2\alpha+\beta = 1$ corresponds to a system with mass-independent mass-to-light ratio (although possible spatial dependence in $\Upsilon$ is still allowed during the Bayesian optimization); $2\alpha+\beta > 1$ indicates that brighter member galaxies exhibit larger $\Upsilon$. 

\section{Cluster Lens Models and Subhalo Dynamics}\label{sec:Cluster_Sample}

Our sample consists of the following cluster lenses with the best-to-date tightly constrained lensing-derived mass models and independently derived copious dynamical data on their cluster member galaxies that we utilize for our analysis.

\subsection{Full cluster sample}\label{ssec:Cluster_Sample_Full}

\begin{table*}[ht]
		\begin{tabular}{ccccccccc}
			\hline Cluster
			& $\<\zspec\>$ & RA [$^\circ$] & Dec [$^\circ$] & $M_{200}$ [M$_\odot$] & $R_{200}$ [Mpc] & $c$ & $\Ngal^{\text{spec}~\cap~\Lenstool}$ &References\\ 	\hline 	\hline 
			Abell 2218 &  0.1710  &  248.9750   &  66.2167  &  $6.80\times 10^{14}$  &  1.72 & 4.96 & 25 & (1,2) \\	
			Abell 383 &  0.1887  &  42.0141  &  -3.5292  &  $6.61\times10^{14}$  &  1.69  & 6.51   & 5 & (3,4)  \\	 
			Abell 963 &  0.2041  &  154.2600   & 39.0484&  $4.07\times10^{14}$  &  1.43  & 7.21  &  2 & (3,5)  \\	 
			Abell 209 &  0.2090  &  22.9703  &  -13.6147  &  $1.40\times10^{15}$  &  2.13  &  3.30 &  9 &  (6,7,8) \\	 
			Abell 2390 &  0.2269 &  328.4060  &  17.6961 & $2.19\times10^{15}$  &  2.47  &  3.24  &  15 &    (3,6,9) \\	 
			MACS J0416 &  0.3972  & 64.0381   & -24.0675   &  $1.53\times10^{15}$  &   2.69 & 2.90  &  66  &  (10,11,12) \\	 
			MACS J1206 &  0.4398  & 181.5506 &   -8.8009  &   $1.37\times10^{15}$  &  1.96  &  5.8 &   152 &  (13,14,15,16) \\ 
			MACS J1149 &  0.5420  &  177.3990  &  22.3979  &  $1.27\times10^{15}$  &  1.84  &  8.8  &  144     &  (12,17) \\	 \hline
		\end{tabular}
	\caption{Galaxy cluster sample and (from left to right) the respective mean spectroscopic redshift $\<\zspec\>$, RA (J2000), Dec (J2000), best-fit NFW virial halo mass $M_{200}$, virial radius $R_{200}$, concentration $c$, number of spectroscopically confirmed and $\Lenstool$-identified member galaxies $\Ngal^{\text{spec}~\cap~\Lenstool}$, and the references: (1) \citet{Mahdavi2001ApJ554L129M}, (2), \citet{Cannon1999MNRAS3029C}, (3) \citet{Newman2013ApJ76524N}, (4) \citet{Geller2014ApJ78352G}, (5) \citet{Rines2016ApJ81963R}, (6) \citet{Koulouridis2021A&A652A12K}, (7) \citet{Annunziatella2016AA585A160A}, (8) \citet{Merten2015ApJ8064M}, (9) \citet{Xu2022AA658A59X}, (10) \citet{Balestra2016ApJS22433B}, (11) \citet{Umetsu2016ApJ821116U}, (12) \citet{Lotz2017ApJ83797L}, (13) \citet{Biviano2023ApJ958148B}, (14) \citet{Biviano:2013eia}, (15) \citet{Umetsu2012ApJ75556U}, (16) \citet{Bergamini2019AA631A130B}, and (17) \citet{Grillo2016ApJ82278G}.}
	\label{tab:Cluster_Properties}
\end{table*}

We list in \tref{tab:Cluster_Properties} the galaxy clusters studied and detail their relevant physical properties below:
\begin{itemize}
	\item Abell 2218 [$z=0.1710$]: We adopted the identification of cluster member galaxies originally cataloged in the \citet{LeBorgne1992AAS9587L} and the \Lenstool-optimized cluster lensing model from \citet{Natarajan:2002cw}\footnote{There exist more recent $\Lenstool$-optimized lens models \citep{Hopwood2010ApJ716L45H, Altieri2010AA518L17A} that are, however, not publicly available.}.
	\item Abell 383 [$z=0.1887$]: This is a cool‐core cluster \citep{Morandi2012MNRAS4213147M} that appears dynamically relaxed and approximately spherical in projection \citep{Newman2013ApJ76524N, Cerini:2022akj, Ueda2020ApJ892100U}. We adopt the member galaxy catalog of \citet{Geller2014ApJ78352G} and \Lenstool-optimized cluster lensing model of \citet{Newman2013ApJ76524N}, where the main cluster halo potential is modeled as a generalized NFW profile $\rho(r) \eee \frac{(3\sigma_\text{Lens}^2/8 G \rs^s)}{\big(\frac{r}{\rs}\big)^{\gamma}\big(1+\frac{r}{\rs}\big)^{3-\gamma}}$. Here, $\gamma\geq 0$ is the central density slope, $\sigma_\text{Lens}$ velocity dispersion, and $\rs$ scale radius.
	\item Abell 963 [$z=0.2041$]: The cluster appears dynamically relaxed and largely spherical in projection \citep{Newman2013ApJ76524N}. We adopt the member galaxy catalog from the SIMBAD database \citep{Wenger2000AAS1439W}\footnote{SIMBAD offers a meta-compilation and as a dynamically updated database, it incorporates all published literature and data from surveys such as SDSS \citep{SDSS2025arXiv250707093S} and DESI \citep{DESI2025arXiv250314745D} as and when these are made publicly available.} and \Lenstool-optimized cluster lensing model of \citet{Newman2013ApJ76524N}, where the main cluster halo potential is modeled as a generalized NFW profile.
	\item Abell 209 [$z=0.2090$]: The cluster appears dynamically relaxed \citep{Gilmour2009MNRAS3921509G, Postman2012ApJS19925P} and largely spherical in projection \citep{Marty2003SPIE4851208M, Smith2005MNRAS359417S}. There exists a CLASH galaxy catalog in the field of Abell 209 but without the membership identification flags of \citet{Annunziatella2016AA585A160A}. We therefore perform an independent membership identification in \aref{app:Member_ID} and adopt the \Lenstool-optimized cluster lensing model of \citet{Smith2005MNRAS359417S}.
	\item Abell 2390 [$z=0.2269$]: This is a strong cool-core cluster \citep{Morandi2007MNRAS3801521M, Sonkamble2015Ap&SS35961S} that appears dynamically relaxed and largely spherical in projection \citep{Newman2013ApJ76524N}. We adopt the member galaxy catalog from the SIMBAD database \citep{Wenger2000AAS1439W} and \Lenstool-optimized cluster lensing model of \citet{Newman2013ApJ76524N}, where the main cluster halo potential is modeled as a generalized NFW profile.
	\item MACS J0416.1-2403 [$z=0.3972$]: The cluster appears dynamically relaxed \citep{Postman2012ApJS19925P}. There exists a CLASH galaxy catalog in the field of MACS~J0416 but without the membership identification flags of \citet{Balestra2016ApJS22433B}. Thus, we perform an independent membership identification in \aref{app:Member_ID} and adopt the \Lenstool-optimized cluster lensing model from the \textit{HST Frontier Fields Initiative} \citep{HSTFrontierFieldsData, Lotz2017ApJ83797L, Natarajan2017MNRAS4681962N}.
	\item MACS J1206.2-0847 [$z=0.4398$]: This is a cool core cluster \citep{Ebeling2009MNRAS3951213E} that appears dynamically relaxed and largely spherical in projection \citep{Postman2012ApJS19925P, Girardi:2015aga}. There exists a CLASH galaxy catalog in the field of MACS~J1206 but without the membership identification flags of \citet{Biviano:2013eia}. We perform an independent membership identification in \aref{app:Member_ID} and adopt the \Lenstool-optimized cluster lensing model of \citet{Caminha2017AA607A93C} that gives $\alpha = 0.28$ and $\beta = 0.64$ for the relationship between mass and light for the subhalo population, in contrast to the scaling relations derived for all others that are consistent with the Faber-Jackson relation.
	\item MACS J1149.6+2223 [$z=0.5420$]: The cluster appears dynamically relaxed \citep{Postman2012ApJS19925P, Finney2018ApJ85958F}. We adopt the member galaxy catalog from the SIMBAD database \citep{Wenger2000AAS1439W} and \Lenstool-optimized cluster lensing model from the \textit{HST Frontier Fields Initiative} \citep{HSTFrontierFieldsData, Lotz2017ApJ83797L, Natarajan2017MNRAS4681962N}. In this work, the NFW parameters of MACS~J1149 in \tref{tab:Cluster_Properties} are obtained by directly fitting the shell-averaged large-scale \Lenstool~halo potentials.  
\end{itemize}

Depending on the depth of the original observations and the number of lensing multiple image families available, these $\Lenstool$-optimized lensing mass maps have included subhalos ranging in number from around $\simeq25$ (for Abell~2218, 383, and 963), 50 (Abell~209 and 2390), and 220 (MACS~J0416, J1206, and J1149) largely within the inner few arcmin region of each cluster ($\lesssim 0.2R_\text{200}$). We next cross-associate these subhalos with published catalogs of spectroscopically confirmed member galaxies. For clusters without a dedicated catalog survey (e.g. Abell~963 and 2390) or with catalogs covering a larger degree-scale field of view extending beyond $R_{200}$ (e.g. Abell~383 and 209), the available member galaxies are rather coarsely sampled within the central one-arcmin region of each cluster, such that many $\Lenstool$ identified subhalos have no matched member galaxies. This low matching efficiency is reflected in the $\Ngal^{\text{spec}~\cap~\Lenstool}$ being noticeably lower than the respective number of subhalos included in the $\Lenstool$ analysis. 

\subsection{The orbital structure of cluster member galaxies}\label{ssec:Cluster_Sample_Full_Anisotropy}

In order to compute the tidal radii, we need to understand the orbital structure of the cluster member galaxies, and having an independent mass estimate from lensing permits this estimate.
The velocity anisotropy $\beta \in(-\infty, 1]$
\begin{align}
    \beta(r) \eee 1 - \frac{\sigma_\text{t}^2(r)}{2\sigma_\text{r}^2(r)},
\end{align}
quantifies the relative ratio between the tangential $\sigma_\text{t}(r)$ and radial $\sigma_\text{r}$(r) components of the velocity dispersion profile; a system with $\beta>0$ ($\beta<0$) consists of a preponderance of more radial (tangential) orbits. The orbits of subhalos and satellite galaxies in group and cluster environments have been demonstrated (statistically) from the SDSS sample \citep{Wojtak2013MNRAS4282407W, Mitra2024MNRAS5333647M} and in cosmological simulations \citep{vandenBosch2019MNRAS4884984V} to be preferentially radial $\beta\sim0.1$\textendash$0.4$. As substructures in more eccentric orbits (smaller pericenter radii) experience a more pronounced loss of tidal mass \citep[e.g.][]{Green2021MNRAS5034075G} and are therefore unlikely to survive. Here we compute and confirm that subhalo orbits are primarily radially anisotropic for the spectroscopically confirmed member galaxies of each cluster. This provides an insight into the impact of tidal stripping for these subhalo populations.

Under the assumption of spherical symmetry, we first numerically invert the projected galaxy number surface density $\Sigmag(R)$ via the Abel inversion \citep[e.g.][]{Binney&Tremaine2008}
\begin{align}\label{eqn:Abel_Inv}
	\nug(r) = -\int_r^\infty \bigg(\frac{d\Sigma_\text{gal}(R)}{dR}\frac{1}{\pi\sqrt{R^2-r^2}}\bigg)dR,
\end{align}
to find the 3D galaxy number density profile $\nug(r)$. The anisotropic Jeans equation yields \citep{Natarajan:1996yt, lokas:hal-00005581, Benatov2006MNRAS370427B}:
\begin{align}\label{eqn:ani_Jeans}
	\frac{d (\nug \sigmar^2)}{dr} + \frac{2\beta \nug\sigmar^2}{r} = - \frac{G M_\text{cluster} \nug}{r^2},
\end{align}
where $\sigmar(r)$ is radial velocity dispersion of member galaxies and $M_\text{cluster}(r)$ is the enclosed total mass of the cluster. Given the line-of-slight velocity dispersion profile $\sigmalos(R)$ computed from each member galaxy catalog (see \aref{app:Member_ID}), the anisotropic Jeans equation can be combined with 
\begin{align}
	\frac{1}{2}[\Sigmag(R) \sigmalos^2(R)] = &\int_R^\infty \frac{r\nug \sigmar^2}{\sqrt{r^2-R^2}}dr \\
    &- R^2 \int_R^\infty \frac{\beta \nug \sigmar^2}{r\sqrt{r^2-R^2}}dr\nonumber,
\end{align}
and reduced to 
\begin{align}
	\nug \sigmar^2 = I_1(r) - I_2(r) + I_3(r) - I_4(r),
\end{align} 
where
\begin{align}
		&I_1 \eee \frac{1}{3}\int_r^{R_\text{max}} \frac{GM_\text{cluster} \nug}{r^2} dr,\\
		&I_2 \eee -\frac{2}{3r^3}\int_0^r GM_\text{cluster} \nug r dr,\nonumber\\
		&I_3 \eee \frac{1}{r^3} \int_0^r R \Sigmag \sigmalos^2 dR,\nonumber\\
		&I_4 \eee \frac{2}{\pi r^3} \int_r^{R_\text{max}} R\Sigmag \sigmalos^2 \bigg[ \frac{r}{\sqrt{R^2-r^2}} -\sin^{-1}\Big(\frac{r}{R}\Big) \bigg]dR,\nonumber
\end{align}
such that $\sigmar^2(r)$ can be evaluated by direct numerical integration; we adopt a fixed integration bound of $R_\text{max} = 10R_{200}$. Lastly, we recover the anisotropy profile by plugging this numerical solution of $\sigmar^2(r)$ into \eref{eqn:ani_Jeans} to obtain:
\begin{align}
	\beta(r) = -\frac{r}{2\nug\sigmar^2} \Big[\frac{GM_\text{cluster} \nug}{r^2} + \frac{d \nug\sigmar^2}{dr}\Big].
\end{align}
Over the radial range populated by the spectroscopically confirmed member galaxies across the entire cluster lens sample, our range of recovered orbital anisotropy $\beta \simeq 0.1$\textendash$0.45$ is consistent with previously published results \citep{Wojtak2013MNRAS4282407W, Mitra2024MNRAS5333647M}.

\section{SUBHALO TIDAL EXTENTS: Analytical Estimates and validation from SIMULATIONS }\label{sec:rt_Estimates}

\subsection{Statistical estimate of tidal truncation radii}\label{ssec:rt_Estimates_Ana}

Consider a member galaxy with a density profile $\rho_{\text{dPIE},i}(r)$, central velocity dispersion $\sigma_{\text{gal},i}$, 3D pericenter radius $r_{\text{per},i}$ and orbital velocity $v_{\text{per},i}$ relative to the galaxy cluster center. For collisionless CDM, the subhalo tidal truncation radius $r_{\text{t},i}$ can be estimated at the ensemble level by using the condition that the mean density enclosed within $r_{\text{t},i}^\text{CDM}$ equals the mean density of the cluster within $r_{\text{per},i}$ \citep{Ghigna1998MNRAS300146G, Taylor2001ApJ559716T}
\begin{align}\label{eqn:CDM_rt}
    \<\rho_{\text{gal},i}(\epsilon_i\times r_{\text{t},i}^\text{CDM})\> \leq \<\rho_\text{cluster}(r_{\text{per},i})\>,
\end{align}
where $\rho_\text{cluster}(r)$ denotes the 3D density profile of the main background potential(s) of each cluster. We note here that this equality is expected to hold for subhalos beyond their first pericenter passages. We emphasize again that this estimate is accurate only in the \textit{statistical sense}, as the details of the tidal stripping process depend sensitively on the individual subhalo infall time \citep{Wu2013ApJ76723W}; orbital parameters \citep[e.g.][]{Green2021MNRAS5034075G}; internal velocity anisotropy \citep{Chiang2024arXiv241103192C}; and the history of the assembly of the host cluster \citep{Hahn2009MNRAS3981742H, Penarrubia2010MNRAS4061290P}. The order-unity prefactor $\epsilon_i$ encapsulates the aforementioned uncertainties in analytical modeling.

In contrast, for SIDM, the additional effect of ram-pressure stripping becomes palpable in subhalos above $\sigma_\text{SIDM}/m_\text{SIDM} \gtrsim 1~\text{cm}^2/\text{g}$ \citep{Nadler:2020ulu} leading to more compact tidal radii. It turns out that in the strongly collisional regime $\gtrsim 10~\text{cm}^2/\text{g}$ that is required for core collapsing subhalos to resolve the GGSL discrepancy, ram pressure stripping becomes the dominant host-subhalo interaction that shapes their morphology and further strips the subhalos to significantly smaller truncation radii \citep{Shirasaki2022MNRAS5164594S}. The analytical estimate, as first worked out by \citet{Furlanetto:2001tw} is given by:
\begin{align}\label{eqn:SIDM_rt}
    \rho_{\text{gal},i}(r_{\text{t},i}^\text{SIDM})\sigma_{\text{gal},i}^2 \leq \rho_\text{cluster}(r_{\text{per},i})v_{\text{per},i}^2
\end{align}
and is based on the physical balance of ram-pressure (left) and external pressure (right) at the new truncation boundary $r_{\text{t},i}^\text{SIDM}$. Once again, this equality holds for subhalos beyond their first pericenter passage at which the mass-loss rate ``spikes,'' a key characteristic of the ram-pressure stripping process \citep[e.g.][]{Vollmer2001ApJ561708V}.

Observationally, one has access only to the projected radius $R_{\text{gal},i}$ and line-of-sight velocity $v_{\text{los},i}\in [0,v_{\text{per},i}]$ of a member galaxy relative to its cluster center. Each \Lenstool-optimized lensing map additionally provides the effective velocity dispersion $\sigma_{\text{dPIE}_i}$ of each substructure, instead of the central velocity dispersion of the associated member galaxy $\sigma_{\text{gal},i}$ that is otherwise observationally inaccessible. The truncation radius estimates can then be roughly recast as\footnote{Here we explicitly account for the ensemble-averaged projection effects in position, which differs slightly from the original analysis by \citet{Natarajan:2002cw}.}
\begin{align}\label{eqn:Obs_rt}
    \begin{aligned}
    &\<\rho_{\text{gal},i}(\epsilon_i r_{\text{t},i}^\text{CDM})\> = \<\rho_\text{cluster}(\sqrt{\frac{3}{2}} R_{\text{gal},i})\>,\\
    &\rho_{\text{gal},i}(r_{\text{t},i}^\text{SIDM})\bigg(\frac{3\sigma_{\text{dPIE},i}^2}{2}\bigg) = \rho_\text{cluster}(\sqrt{\frac{3}{2}} R_{\text{gal},i}) v_{\text{los},i}^2,
    \end{aligned}
\end{align}
where now $r_{\text{t},i}^\text{SIDM}$ is a conservative upper bound due to velocity projection $v_{\text{los},i}$. It is required that $v_{\text{los},i}$ is not greater than the unprojected instantaneous subhalo orbital velocity, which in turn is not greater than $v_{\text{per},i}$.

Empirically, \citet{Ghigna1998MNRAS300146G} found that the $r_{\text{t},i}^\text{CDM}$ estimate with a typical scatter $\epsilon_i \simeq 0.5$\textendash$2$ agreed reasonably well with the definition of subhalo tidal radius by \citet{King.62}, across all projected radius bins in their cluster-scale cosmological simulation. However, their main cluster comprised only $6\times 10^5$ particles within the virial radius, and all substructures with $\geq32$ particles were considered resolved in their subhalo sample, which are bound to suffer from the ``overmerging'' numerical artifacts \citep[e.g.][]{Penarrubia2010MNRAS4061290P, vandenBosch2018MNRAS4754066V, vandenBosch2018MNRAS4743043V, Benson2022MNRAS5171398B, Chiang2025arXiv251026901C}. Furthermore, there exist several other commonly adopted definitions of tidal radius in the literature that all assume different physical simplifications (see \citet{Tollet.etal.17} and \citet{vandenBosch2018MNRAS4743043V} for detailed discussions).

To assess numerical robustness and avoid complications arising from different definitions of analytical tidal radius, we anchor density-based CDM estimates \eref{eqn:Obs_rt} to lensing-based tidal truncation radii $r^\text{dPIE}_{\text{t},i}$ described above. We then consistently and directly infer the calibration factor $\epsilon_i$ and subhalo-to-subhalo variance thereof from the Illustris-TNG simulations. Such population-level calibration to $r^\text{dPIE}_{\text{t},i}$ defined in \eref{eqn:rho_dPIE} also allows us to make direct comparison between our CDM estimates and observationally inferred values for tidal truncation. In \sref{ssec:rt_Estimates_Ana_Sim}, we compute the distribution of $\epsilon_i$ for simulated CDM subhalos in the redshift- and mass-matched cluster analogs from the Illustris-TNG simulation. We do not repeat the same numerical exercise for SIDM, as substructures in existing SIDM cosmological simulations still suffer from particle resolution and numerical convergence issues \citep{Mace2024arXiv240201604M, Palubski2024JCAP09074P,Fischer2024}\footnote{The SIDM Concerto \citep{Nadler2025arXiv250310748N} achieves more conservative particle resolution among existing cosmological simulations. However, their cluster host with $M_{200} = 1.6\times10^{14}$~M$_\odot$ is still quite less massive relative to our sample (\tref{tab:Cluster_Properties}), and $\sigma_\text{SIDM}/m_\text{SIDM} \simeq 0.1~\text{cm}^2/\text{g}$ on cluster scales adopted therein does not produce any core-collapsing subhalos.}. We instead adopt the conservative upper bound $r_{\text{t},i}^\text{SIDM} \eee \min(r_{\text{t},i}^\text{CDM}, r_{\text{t},i}^\text{SIDM})$ for SIDM estimates in \sref{sec:rt} as appropriate cosmological SIDM simulations are currently unavailable.

Lastly, in the original derivation of \eref{eqn:CDM_rt} by \citet{Taylor2001ApJ559716T}, the background host mass distribution and potential were both shell-averaged. In comparison, \Lenstool-optimized cluster mass models do allow multiple large-scale cluster background potentials with non-zero ellipticities. Following the methodology of \citet{Natarajan:2002cw}, we first compute the projected Center of Mass (CoM) for each cluster as the arithmetic mean of CoM of all large-scale potentials, weighted by the respective total mass within $r_\text{t}$. The cluster projected density profile $\Sigma_\text{cluster}(R)$ is the linear superposition of all large-scale clumps, shell-averaged with respect to the common CoM. The final 3D cluster density profile $\rho_\text{cluster}(r)$ is obtained by deprojecting $\Sigma_\text{cluster}(R)$ via \eref{eqn:Abel_Inv}. For clusters in our sample (\tref{tab:Cluster_Properties}) whose large-scale cluster potentials have published best-fit NFW density profiles \citep{Navarro:1996gj}, we verify in \aref{app:Lenstool_vs_NFW} that the CDM estimates from both background potentials derived from \Lenstool~and NFW are consistent.

\subsection{Cross validation of CDM tidal radii estimates with clusters in the Illustris-TNG simulation}\label{ssec:rt_Estimates_Ana_Sim}

We evaluated the tidal radius estimate \eref{eqn:Obs_rt} in the Illustris-TNG $\Lambda$CDM cosmological magnetohydrodynamical simulation suite \citep{Nelson2019ComAC62N}, performed with the
moving-mesh code \textsc{Arepo} \citep{Springel2010MNRAS401791S}. We start by selecting the closest redshift- and mass-matched massive cluster analogs as to our observed lens sample  \tref{tab:Cluster_Properties}. Here, we focus on a new simulation set, the TNG-Cluster\footnote{\url{https://www.tng-project.org/cluster/}}~\citep{Nelson2024AA686A157N} that has a simulation volume of size $1003.8$~Mpc with periodic boundary conditions; wherein the dark matter (baryon) particles have a fixed mass resolution of $6.1\times10^7$ ($1.2\times10^7$)~M$_\odot$ and gravitational softening length of $1.48$~kpc. We first identify the corresponding closest redshift-matched data outputs of our sample: Abell~2218 (snap86), Abell~383 (snap85), Abell~963 (snap84), Abell~209 (snap83), Abell~2390 (snap82), MACS~J0416 (snap72), MACS~J1206 (snap70), and MACS~J1149 (snap65). And in each snapshot, we select five best mass-matched cluster systems for further study, with a $\Delta M_{200}/M_{200} = 0.00038$\textendash$0.28$ in our final sample of cluster analogs.

For each cluster analog, we extract the particle data of the host halo and all subhalos in the mass range $M^\text{subhalo} = 10^{10.5\text{\textendash}12.5}$~M$_\odot$ (equivalent to $\sim5\times 10^{2\text{\textendash}4}$ dark matter particles per system). Although these substructures are still expected to suffer from the numerical overmerging issue \citep{vandenBosch2018MNRAS4754066V, Chiang2025arXiv251026901C}, they still represent a significant improvement compared to the subhalo sample analyzed in \citet{Ghigna1998MNRAS300146G}. The shell-averaged density profile is then computed for each (sub)halo with respect to the respective CoM.

For each selected subhalo, we first compute $r_{\text{t},i}^\text{CDM}$ from \eref{eqn:Obs_rt} by setting $\sqrt{3/2} R_{\text{gal},i}$ as its instantaneous 3D distance from the host CoM with $\epsilon_i \eee 1$ (i.e., perfect equivalence with $r_{\text{t},i}^\text{dPIE}$). Next, in lens-based inference, only the truncated radius $r_{\text{t},i}^\text{dPIE}$ in $\rho_\text{dPIE}(r)$ is relevant here. The subhalo potentials in the \Lenstool-optimized lensing cluster models of our sample (\tref{tab:Cluster_Properties}) all feature sub-kpc core radii that are two to three orders of magnitude smaller than inferred truncation radii $r_{\text{core},i} \ll r_{\text{t},i}$ and well below the TNG-Cluster resolution scale. Instead of fitting all three parameters in $\rho_0$, which inevitably mix in resolution-limited numerical artifacts, we use the empirical fact that $r_{\text{core},i} \ll 0.1r_{\text{max},i}$ to simplify \eref{eqn:rho_dPIE} in the limit $\rho_\text{dPIE}(r\gg r_{\text{core},i}) = \frac{\rho_{\text{eff},i}}{r^2[1+(r/r_{\text{t},i}^\text{dPIE})^2]}$, where $r_{\text{max},i}$ denotes the maximal radial extend of the subhalo in question. We then determine the prefactor $\rho_{\text{eff},i}$ (irrelevant to the present discussion) and $r_{\text{t},i}^\text{dPIE}$ by matching the total integrated mass and performing a least squares fitting (in log-log space) to each shell-averaged subhalo density profile between $0.1$\textendash$1r_{\text{max},i}$. This approach is numerically stable in extracting $r_{\text{t},i}^\text{dPIE}$ and unaffected by uncertainties at small radii.
\begin{figure}
	\includegraphics[width=\linewidth]{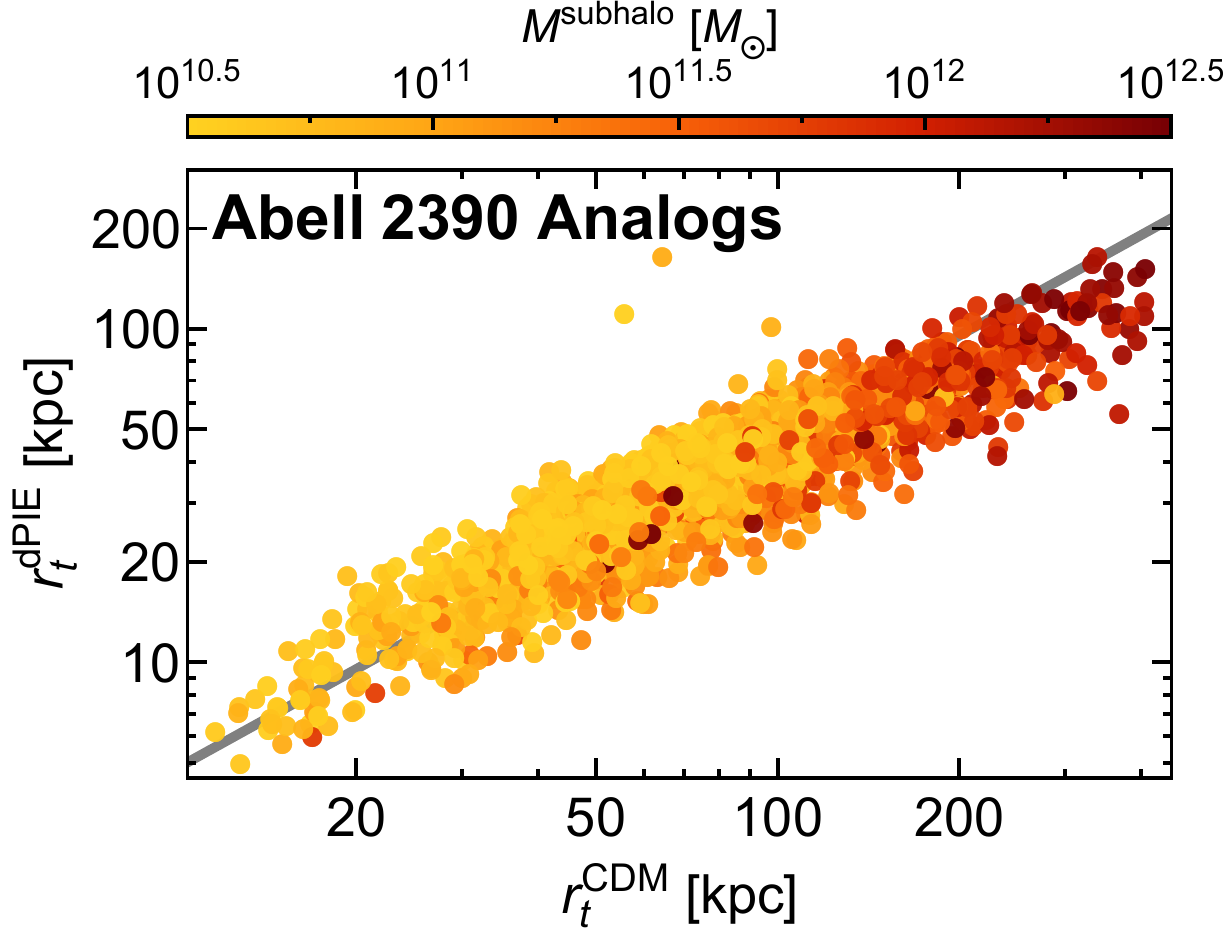}
	\caption{Comparison of density-based $r^\text{CDM}_\text{t}$ and lensing-based $r^\text{dPIE}_\text{t}$ tidal truncation radius estimates for subhalos with masses $10^{10.5\text{\textendash}12.5}$~M$_\odot$ (color-coded) in simulated Abell~2390 cluster analogs. The median $\tilde{\epsilon_i} = 0.477$ (gray line) is statistically representative across the entire sample of 2860 subhalos.}
	\label{fig:TNG_Cluster-2}
\end{figure}

To quantify the statistical distribution of $\epsilon_i$, \fref{fig:TNG_Cluster-2} compares the individual estimates $r_{\text{t},i}^\text{CDM}$ ($x$-axis) and $r_{\text{t},i}^\text{dPIE}$ ($y$-axis) for subhalos within $M^\text{subhalo} = 10^{10.5\text{\textendash}12.5}$~M$_\odot$ in the Abell~2390 cluster analogs. The median $\tilde{\epsilon_i} = 0.477$ (gray line) captures the trend averaged over the ensemble for the range of subhalo masses and sizes examined here. Most importantly, we note the general statistical consistency in the median relation against subhalo properties, except for a minor systematic deviation above $r_\text{t}^\text{dPIE} \gtrsim 100$~kpc that could be interpreted as suggestive of the existence of resolution-limited numerical artifacts; we return to this point in \sref{sec:rt_Uncertainties}.

\begin{figure}
	\includegraphics[width=\linewidth]{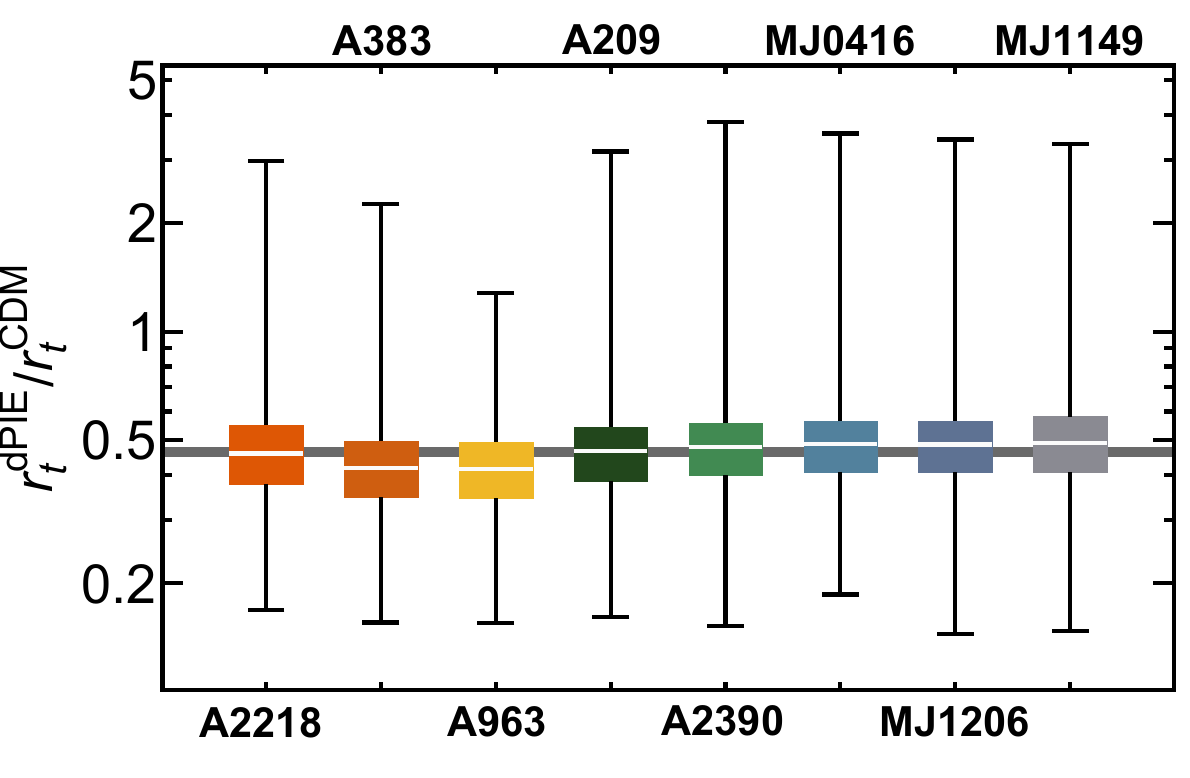}
	\caption{Boxplot displaying the $r^\text{dPIE}_\text{t}/r^\text{CDM}_\text{t}$ distributions, which exhibit high consistency across all redshift- and mass-matched TNG-Cluster analogs of our sample (\tref{tab:Cluster_Properties}), arranged in increasing redshifts from left to right. From a final sample of 13714 subhalos in all analogs, we infer a median of $\tilde{\epsilon_i} = 0.470$ (gray line) and interquartile range of 0.391\textendash0.553. The central 97\% entries lie within 0.248\textendash1.18.}
	\label{fig:TNG_Cluster-3}
\end{figure}

In general, we find that the median and typical subhalo-to-subhalo variance derived \fref{fig:TNG_Cluster-3} in the calibration factor $\epsilon_i$ are consistently similar among all cluster analogs matched to redshift and mass. The factor four to five range in the central 97\% distributions is consistent with the typical scatter reported in \citet{Ghigna1998MNRAS300146G} and is expected due to sizable subhalo-to-subhalo variation in their physical properties (e.g., formation time, accretion time, orbits, velocity anisotropies, and ellipticities). This order-unity scatter also underscores the prudent note that  the tidal truncation estimates derived here, \eref{eqn:Obs_rt}, should be interpreted only on an \textit{ensemble} level as done in \sref{sec:rt}, and should not be used to derive dark matter constraints on an object-by-object basis. In the subsequent analysis, we therefore quote the median, interquartile range, and central 97\% range of $\epsilon_i$ from \fref{fig:TNG_Cluster-3} as our calibration factor and physical variances thereof.

\section{CDM vs. SIDM: Lensing-inferred constraints from Tidal Truncation Properties}\label{sec:rt}

\begin{figure*}
	\includegraphics[width=\linewidth]{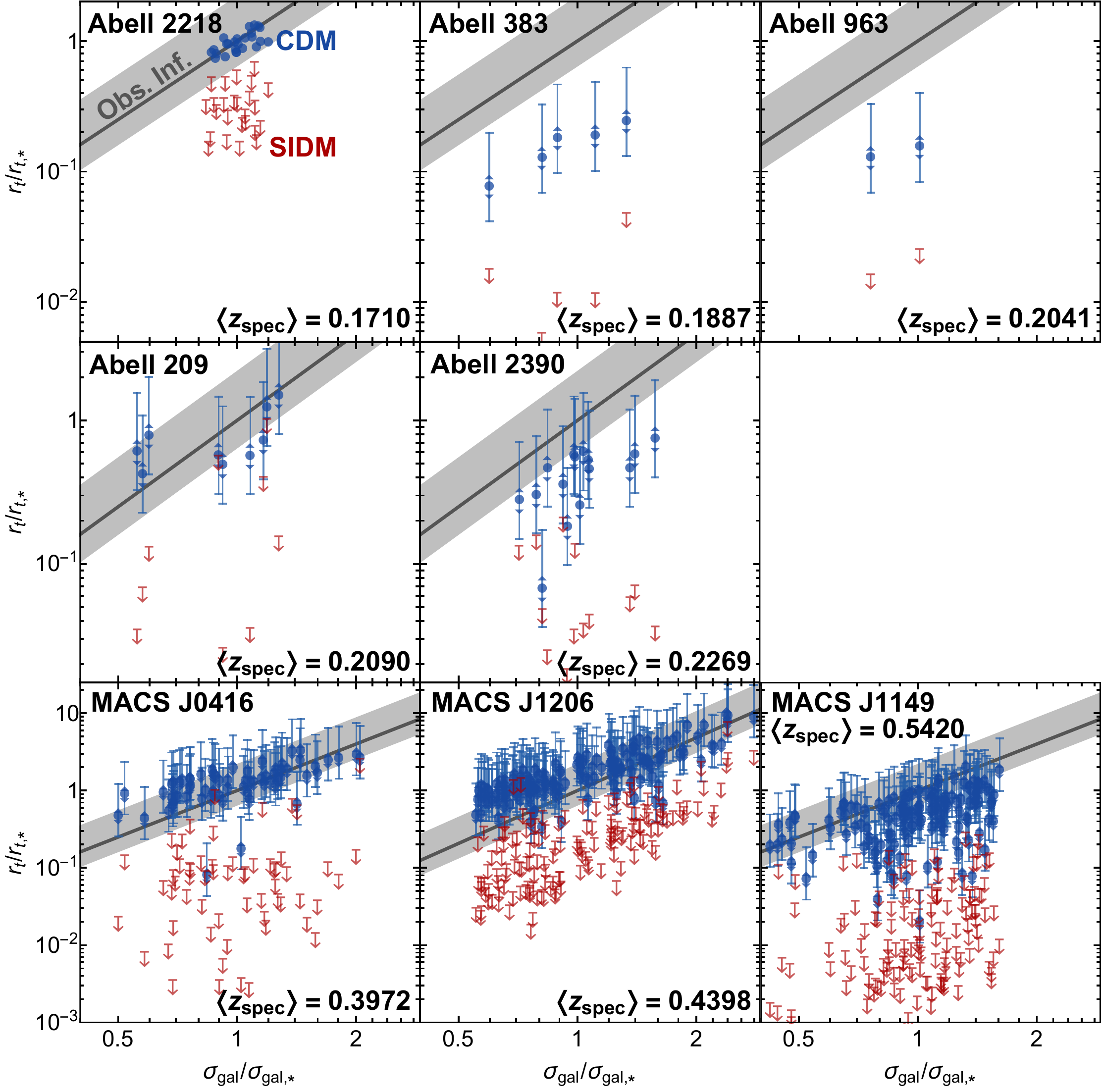}
	\caption{Tidal truncation radii of cluster subhalos derived from lensing-based observational inference (gray curves; gray-shading indicates conservative $5\sigma$), CDM estimates (blue circles with $\tilde{\epsilon_i} = 0.470$; arrowheads (error bars) denote the central 50\% (97\%) range of subhalo-to-subhalo variance inferred from TNG-Cluster in \fref{fig:TNG_Cluster-3}), and SIDM estimates (red; conservative upper bounds). Data points of Abell~2218 are quoted directly from \citet{Natarajan:2002cw}. \textit{On a population level, CDM is consistent with, while SIDM is ruled out with high statistical significance across, the entire cluster sample.} Modulo detailed cluster assembly history, the overall distribution of CDM-predicted tidal radii steadily shifts downward with decreasing redshifts (or increasing cosmic age from $t_\text{Universe} = 8.1$~Gyr to $11.4$~Gyr), indicative of continuous subhalo tidal striping. The consistent one-to-two order-of-magnitude discrepancy in SIDM predictions below the observational inference cannot be reconciled with uncertainties in either observational inference or subhalo-to-subhalo variance in $\epsilon_i$, ruling out dark matter collisionality in cluster subhalos of the mass range $M^\text{subhalo}_\text{dPIE} \simeq 5\times 10^{9\text{\textendash}12}$~M$_\odot$.}
	\label{fig:CDM_SIDM}
\end{figure*}

Collisionless CDM and (strongly) collisional SIDM subahlos are predicted to exhibit starkly different tidal truncation extents (\sref{ssec:rt_Estimates_Ana}), which itself poses a well-defined self-consistency test of dark-matter models when compared against observations (\sref{ssec:Cluster_DM_rt}). \fref{fig:CDM_SIDM} compares the lensing-inferred (gray curves), CDM-predicted (blue), and SIDM-predicted (red) tidal truncation radii of cluster member galaxies that are both confirmed and identified spectroscopically in our optimized mass models $\Lenstool$. For the case of Abell~2218, we plot the data points directly from \citet{Natarajan:2002cw}. We additionally adopt very conservative $5\sigma$ bounds of inference uncertainty (gray shading) quoted in \citet{Natarajan:2002cw} for all clusters.

As evident in \fref{fig:CDM_SIDM}, the high statistical consistency between the empirically inferred and CDM predictions across all eight clusters indicates that collisionless tidal stripping is the dominant and probably the only dynamical process that operates to produce the observed subhalo tidal extents. Indeed, in the scenario where the collisional nature of dark matter particles becomes dominant, the significantly compact subhalo tidal radii are strongly discrepant with the lensing inference across all eight independent massive cluster environments $M_{200} = 0.41$\textendash$2.2\times10^{15}$~M$_\odot$ and redshift range $\<z_\text{spec} \> \simeq 0.17$\textendash$0.54$ probed in our sample. 

Importantly, these subhalos associated with the spectroscopically confirmed member galaxies span orders of magnitude in their lensing-inferred total mass ranging over $M^\text{subhalo}_\text{dPIE} \simeq 5\times 10^{9\text{\textendash}12}$~M$_\odot$ (or equivalently $\sigma_\text{gal} \simeq 40$\textendash$300$~km~s$^{-1}$). The persistently strong discrepancy with SIDM predictions well down to the dwarf scale $\sigma_\text{gal} \simeq 40$~km~s$^{-1}$ also strongly disfavors current models with velocity-dependent cross sections \citep[e.g.][]{Feng:2009mn, Buckley:2009in, Tsai:2020vpi} that are invoked to explain the phenomenology on dwarf galaxy mass scales $\sigma_\text{gal} \lesssim 80$~km~s$^{-1}$ while evading stringent cross section constraints on cluster scales $\sigma_\text{gal} \sim 1500$~km~s$^{-1}$ \citep[e.g.][]{Loeb:2010gj, Tulin:2012wi, Chu:2018fzy, Chu:2018faw}. For the latter class of models, one would naïvely expect that observationally inferred tidal truncation radii ought to be correlated with the (projected) subhalo orbital velocity, a feature that we clearly do not observe for any of the best-fit cluster lensing models for the sample studied here. Our results firmly establish that dark matter self-interaction has to be essentially negligible, if at all present, in cluster environments down to mass scales of $\sim 5\times 10^{9}$~M$_\odot$, analogous to those of individual dwarf galaxies in the field, that have been used to make the case for SIDM and its variants \citep[e.g.][]{Tulin:2017ara}. 

\section{Sources of uncertainty in our analysis}\label{sec:rt_Uncertainties}

In this section, we carefully assess the potential sources of uncertainties in our  determination of subhalo tidal extents:
\begin{itemize}
    \item Observational Inference: (1) \textit{Observational systematics} is largely mitigated with diverse data collection and reduction pipelines. Our sample comprises independent observations of eight galaxy clusters over 25 years with data reduced by several groups \citep[e.g.][]{LeBorgne1992AAS9587L, Smith2005MNRAS359417S, Newman2013ApJ76524N, Lotz2017ApJ83797L, Caminha2017AA600A90C}; therefore, it is extremely unlikely that identical systematics persists across our entire cluster sample. (2) \textit{Lensing mass map construction} adopted in this work, $\Lenstool$, has been extensively tested and verified to produce convergent results against other alternative construction techniques \citep[e.g.][]{Lotz2017ApJ83797L, Caminha2017AA607A93C, Caminha2019A&A632A36C, Natarajan2024SSRv22019N}. In particular, the lens modeling comparison project by \citet{Meneghetti2017MNRAS4723177M} reported that $\Lenstool$ robustly recovers the properties of subhalos, like their mass, in an unbiased fashion. Detailed comparison with other lens mass reconstruction methods showed that  all parametric methods are in excellent agreement for integrated quantities like total mass partitioned into smaller scale subhalos, the most relevant metric for the analysis presented in this work. Non-parametric and hybrid lens mass reconstruction are not well suited for direct comparison with cosmological simulations in which bound structures are conceptualized as halos and subhalos. To guard against the detection of spurious substructures, which is particularly prevalent on the low-mass end \citep{Ephremidze2025MNRAS5422610E}, here we select only subhalos that are cross-verified to host spectroscopically confirmed member galaxies. (3) \textit{Empirical scaling relations} of the subhalo and member galaxy properties, \eref{eqn:E_Scaling_Relations}, are observationally motivated \citep{Faber1976ApJ204668F, Natarajan:2002cw}, optimized in a highly flexible multiscale Bayesian framework \citep{Kneib1996ApJ471643K, Natarajan1997MNRAS287833N}, and verified for their robustness in matching observables \citep{Richard2010MNRAS404325R, Eichner2013ApJ774124E, Desprez2018MNRAS4792630D}. We hence conclude that our observational inferences of subhalo tidal extents appear to be robust against these known uncertainties.
    \item CDM predictions: Accurate predictions of tidally evolved subhalo properties are not possible without detailed knowledge of subhalo accretion history, orbital parameters, initial internal velocity anisotorpy \citep[e.g.][]{Ogiya2019MNRAS485189O, Errani2020MNRAS4914591E, Chiang2024arXiv241103192C}. Here, we seek to estimate distributions of subhalo tidal truncation radii that are accurate at an ensemble level (\sref{ssec:rt_Estimates_Ana}) by calibrating and cross-validating against state-of-the-art large-box cosmological simulations (\sref{ssec:rt_Estimates_Ana_Sim}). The main source of uncertainty in the simulations arises from the fact that these TNG-Cluster subhalos can suffer from mass-resolution-limited numerical artifacts that artificially enhance subhalo tidal mass loss  \citep{vandenBosch2018MNRAS4754066V, Chiang2025arXiv251026901C}. In particular, their adopted softening length of 1.48~kpc is (significantly) larger than the lensing-inferred core radius associated with most of the member galaxies in our sample. One can be reasonably concerned about the numerical robustness of the values of the tidal radii of simulated subhalos. Despite this, we emphasize that we perform the tidal radius calibration factors $\epsilon_i$ on an object-by-object basis (\fref{fig:TNG_Cluster-2}). Namely, this is a calibration between two slightly different definitions of tidal truncation radius, which should not be significantly impacted even when a subhalo is inadequately force-resolved and if it undergoes artificial mass loss. Furthermore, during the optimization of each lens model (\sref{ssec:Cluster_DM_rt}), the core and tidal truncation radii are independently normalized, as they are independent parameters for the adopted density profile, therefore, we also decouple these two characteristic scales in our calibration process. Therefore, we expect the present analysis to be largely robust against numerical resolution-related issues. We leave it for future work to quantify this convergence with even higher-resolution cosmological simulations.
    \item SIDM predictions: Massive cluster-scale SIDM cosmological simulations with mass resolution comparable to or higher than TNG-Cluster for numerical convergence \citep{Mace2024arXiv240201604M, Palubski2024JCAP09074P} are currently not available. Therefore, we resort only to analytical estimates of the tidal extents \citep{Furlanetto:2001tw} that are based on well-understood ram pressure stripping of a collisional fluid that have been numerically validated \citep{Kim2009ApJ7031278K, Bernal2013ApJ77572B, Morton2021arXiv210315848M}. We leave the detailed comparison between analytical and numerical estimates of SIDM subhalo tidal extents in cluster environments for future work. 
\end{itemize}

\section{Implications of our results and other cluster-scale tests}\label{ssec:rt_Implications}

In this section, we discuss the implications of our analysis for the collisional nature of dark matter and place our results in the context of relevant current literature.

Numerous independent studies of observed properties of massive galaxy clusters have been used to probe dark matter and set limits on its collisionality, including studies of the geometry of strong gravitationally lensed arcs \citep{Meneghetti2001MNRAS325435M}; lensing-inferred density profiles \citep{Kaplinghat2016PhRvL116d1302K, Elbert2018ApJ853109E, Sagunski2021JCAP01024S, Andrade2022MNRAS51054A}; cluster collisions and mergers \citep{Markevitch:2003at, Randall:2008ppe, Harvey:2015hha}; offset in the cluster core \citep{Massey2018MNRAS477669M} or brightest cluster galaxies \citep{Lauer2014ApJ79782L, Kim2017MNRAS4691414K, Harvey2019MNRAS4881572H}. All of these diagnostics still tend to favor dark matter being collisionless with tight constraints on the velocity independent self-interaction cross section to be $\sigma_\text{SIDM}/m_\text{SIDM} \lesssim 0.1$\textendash$1$~cm$^2$~g$^{-1}$. 

Performing a stacking analysis on cluster scales, \citet{Banerjee2020JCAP02024B} and \citet{Bhattacharyya2022ApJ93230B} attempted to constrain the average density profiles by comparing cluster weak lensing data with simulated analogs in SIDM cosmological simulations and derived constraints on the self-interaction cross-section of $\sigma_\text{SIDM}/m_\text{SIDM} \lesssim 2$~cm$^2$~g$^{-1}$. However, the limited numerical resolution (with only $\sim10^{5\text{\textendash}6}$ particles in total enclosed within the entire virial radius of each cluster) implies that these results are subject to known numerical artifacts \citep{vandenBosch2018MNRAS4754066V, Mace2024arXiv240201604M, Palubski2024JCAP09074P, Chiang2025arXiv251026901C}. 

In our work, we have focused on smaller scales examining the outer tidal extents of cluster member subhalos associated with the spectroscopically confirmed galaxies, and we find that current lensing observations on these scales are consistent with CDM. As noted in \sref{sec:rt}, our analysis excludes the presence of dark-matter self-interaction down to dwarf galaxy scales in clusters, consistent with the recent constraint $\sigma_\text{SIDM}/m_\text{SIDM} \lesssim 0.2$~cm$^2$~g$^{-1}$ at $\sigma_\text{gal} \sim 20$~km~s$^{-1}$ derived from the kinematics of Milky Way satellite dwarf galaxies \citep{Ando2025arXiv250313650A}. 

Although our analysis convincingly excludes SIDM in the strongly collisional regime, we do not report a specific exclusion bound for the following reasons. First, the analytical estimate \eref{eqn:SIDM_rt} yields tidal truncation radii of SIDM subhalos in the strongly collisional limit. Since subhalo tidal evolution is a highly non-linear process, one cannot simply interpolate between \eref{eqn:SIDM_rt} and the CDM prediction \eref{eqn:CDM_rt} to reliably predict the expected subhalo truncation radii as a function of $\sigma_\text{SIDM}/m_\text{SIDM}$. Second, \eref{eqn:SIDM_rt} is actually determined by the 3D subhalo orbital velocity at peri-center passages, while observationally we only have access to the subhalo's instantaneous projected velocity. We therefore underestimate the statistical discrepancy of SIDM predictions against observations (\fref{fig:CDM_SIDM}). Third, the factor four to five subhalo-to-subhalo scatter in the CDM tidal truncation calibration factor, $\epsilon_i$ (\fref{fig:TNG_Cluster-3}), also implies that one cannot derive a meaningful exclusion bound on an object-by-object basis. Instead, an ensemble-level comparison between simulations and observational inferences is what can be meaningfully and is required to marginalize over both the projection effects and subhalo-to-subhalo variance. For future work, this cluster-scale diagnostic can be particularly instrumental in constraining SIDM models with velocity-dependent cross-sections \citep[e.g.][]{Nadler2025arXiv250310748N}, given the mass range of $M^\text{subhalo}_\text{dPIE} \simeq 5\times 10^{9\text{\textendash}12}$~M$_\odot$ probed.

\section{Summary and Conclusions}\label{sec:Conclusions}

We present the first constraints on dark matter microphysics from  the outer truncation radii of subhalos in massive galaxy clusters. Examining the outer subhalo structure, we find that the lensing-inferred tidal truncation radii are in excellent agreement with collisionless CDM, ruling out significant self-interaction effects at cluster-member scales.

We have presented a detailed analysis of the derived tidal extents of subhalos associated with the spectroscopically confirmed cluster member galaxies in eight independent massive cluster environments $M_{200} = 0.41$\textendash$2.2\times10^{15}$~M$_\odot$, spanning the redshift range $\<z_\text{spec} \> \simeq 0.17$\textendash$0.54$. The inferred subhalo tidal truncation radii are extracted from the existing \Lenstool~cluster mass models constructed from flexible Bayesian optimization, combining strong and weak lensing data simultaneously. By comparing the robustly inferred tidal radii distributions with predictions from collisionless CDM and collisional SIDM paradigms, we clearly demonstrate in \fref{fig:CDM_SIDM} that dark matter self-interactions are negligible, if at all present, in cluster environments on mass scales of $5 \times 10^{9\text{\textendash}12}$~M$_\odot$. Our results have important implications for alternate dark matter models\textemdash CDM-like models with negligible self-interaction are preferred overall compared to either constant or velocity dependent interaction cross-sections. Future data releases of large spectroscopic surveys such as DESI \citep{DESI2025arXiv250314745D} and cluster lensing maps sharpened by new measurements from 
e.g. JWST \citep{Acebron2025AA699A101A} or Euclid \citep{Euclid:2025ado} observations are expected to reveal even more low-mass substructures and permit further mapping of the distribution of tidal truncation radii down to smaller $\sigma_\text{gal}$. The metrics explored in this analysis offer a promising path forward for discriminating between CDM and SIDM models.

\section*{Acknowledgements}
We thank Andrew Newman for providing the optimized $\Lenstool$ models of Abell 383, Abell 963, and Abell 2390. We also thank Graham Smith for sharing the $\Lenstool$ model of Abell 209. We additionally thank Amata Mercurio and Andrea Biviano for providing the rest-frame $\sigmalos$ profile of Abell 209. The material presented in this paper is based upon work supported by NASA under award No. 80NSSC25K0311 to I.D. under the NASA FINESST program. P.N. gratefully acknowledges funding from the Department of Energy grant DE-SC001766 and support from the John Templeton Foundation via grant 126613 for this work.

\section*{Data Availability}

The $\Lenstool$-optimized cluster lensing models for the clusters MACS J0416.1-2403, MACS J1206.2-0847, and MACS J1149.6+2223 are publicly available at \url{https://archive.stsci.edu/prepds/frontier/lensmodels/}. The rest of the data used in the workthis analysis will be shared upon reasonable request to the corresponding author.

\newpage
\appendix

\section{Cluster membership identification}\label{app:Member_ID}

\begin{figure}
	\includegraphics[width=0.5\linewidth]{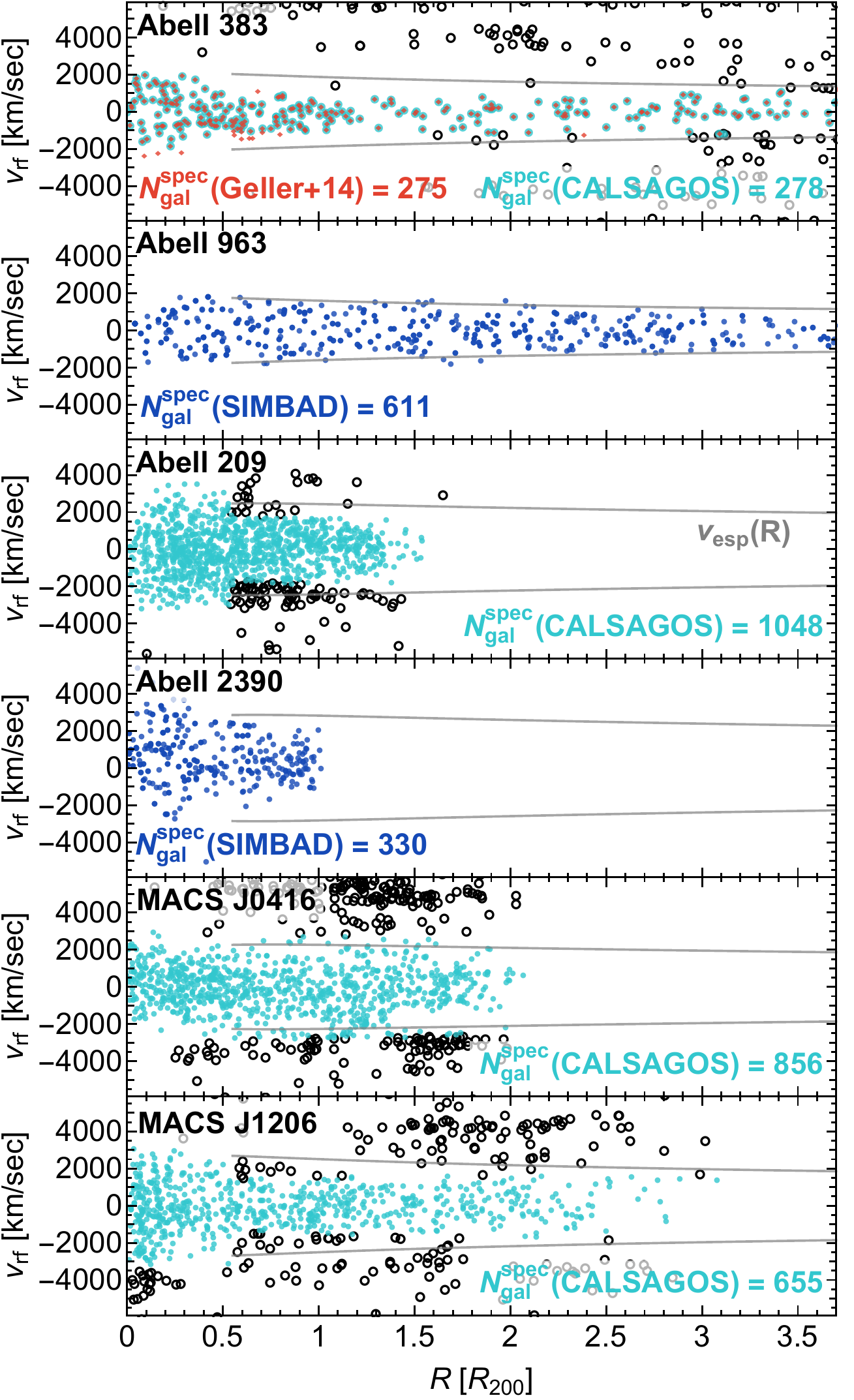}
	\includegraphics[width=0.5\linewidth]{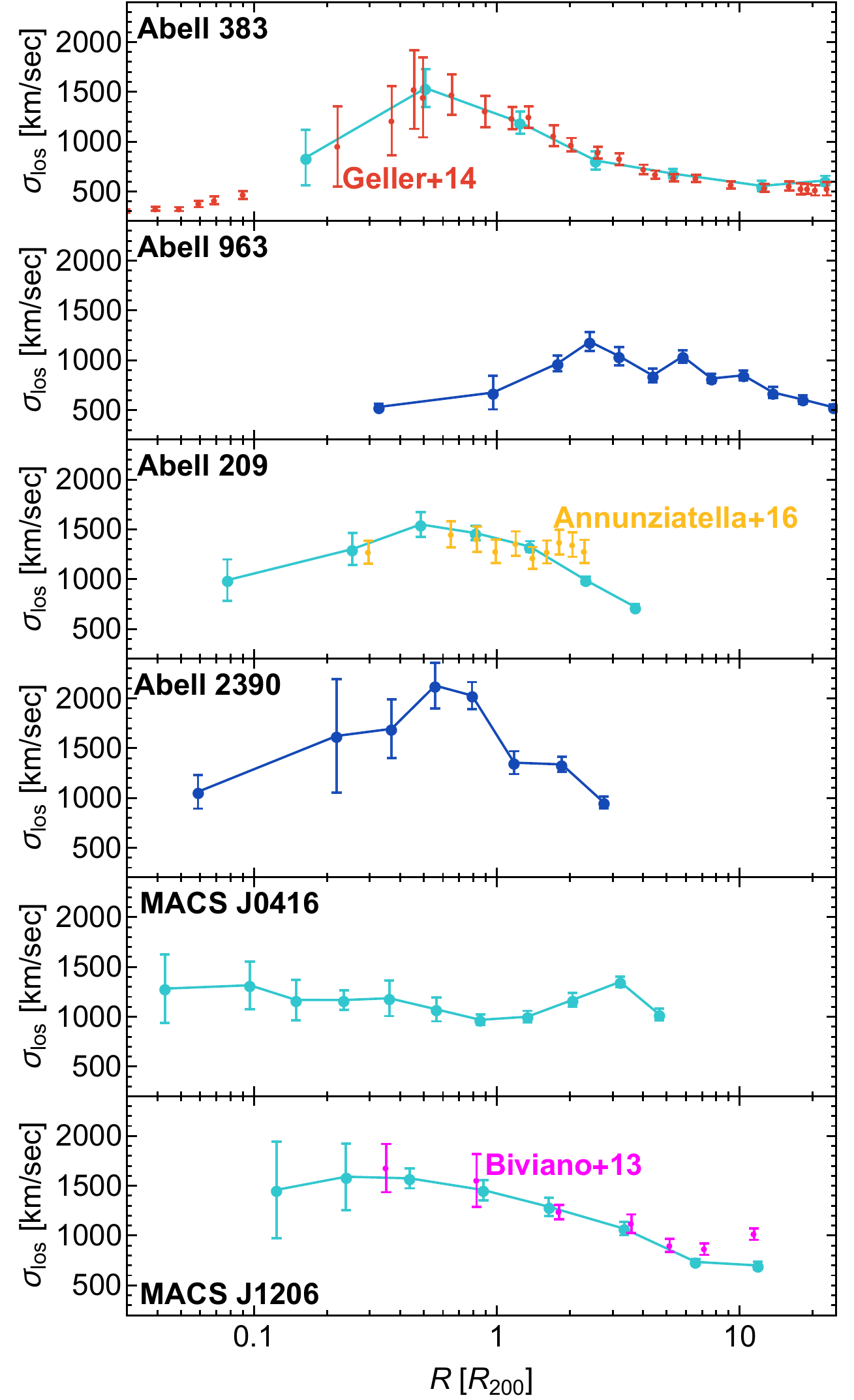}
    \caption{\textit{Left:} Projected rest-frame velocity $v_\text{rf}$ of cluster galaxy sources. We compare the galaxy membership identification from our \texttt{CALSAGOS}-based procedure (cyan), SIMBAD database \citep{Wenger2000AAS1439W} (dark blue), or \citet{Geller2014ApJ78352G} (red; for Abell~275); non-members are marked with open black circles. Gray curves denote the escape velocity curve $v_\text{esp}(R)$ inferred from the lensing-constrained NFW mass profile. \textit{Right:} Line-of-sight velocity dispersion profiles of each member galaxy sample. Our independent membership identification routine yields $\sigma_\text{los}(R)$ in great agreement with published data from \citet{Geller2014ApJ78352G} (red; for Abell~275),  \citet{Annunziatella2016AA585A160A} (yellow; for Abell~209),  \citet{Biviano:2013eia} (magenta; for MACS~J1206).}
	\label{fig:Cluster_Membership_Identification}
\end{figure}

Three clusters in our sample\textemdash Abell~209, MACS~J0416, and MACS~J1206 (see \sref{ssec:Cluster_Sample_Full})\textemdash have public spectroscopic catalogs that contain all observed sources within the field of view but lack cluster membership identification. In this Appendix, we detail the galaxy membership identification process and cross-validate our approach with available literature results. In the pre-processing step, for all sources lying outside a projected distance $R>0.55 R_{200}$ from the cluster center, we first conservatively exclude all sources with a projected rest-frame velocity $\vrf$ exceeding 1.5$\vesp(R)$ from further membership consideration \citep[e.g.][]{Geller2014ApJ78352G, Biviano2021A&A650A105B}. Here, $\vesp(R)$ denotes the escape velocity at a projected radius $R$ inferred from each best-fit cluster large-scale NFW potential listed in \tref{tab:Cluster_Properties}. 

Next, we determine the full membership association of each pre-processed spectroscopic catalog using the open-source Python package \texttt{CALSAGOS} \citep{Olave-Rojas2023MNRAS5194171O} that employs clustering algorithms \citep{OlaveRojas2018MNRAS4792328O} to identify substructures, galaxy groups, and member galaxy associations from spectroscopic and photometric catalogs of galaxy clusters. Specifically, we use the \texttt{CLUMBERI} module therein that decomposes member galaxy clustering into a collection of 3D Gaussian distributions via the method of Gaussian mixture models \citep{Muratov2010ApJ7181266M}. The iterative optimization is based on the Bayesian Information Criterion \citep{Schwarz1978AnSta6461S} and outputs bootstrap estimation of velocity dispersion uncertainty once the best decomposition model is identified. The left panels of \fref{fig:Cluster_Membership_Identification} show the galaxies identified as members (colored dots) or non-members (open black circles). For Abell~383 with existing member associations from \citet{Geller2014ApJ78352G} (red), our analysis scheme yields a nearly identical membership identification out to $\simeq 3.5 R_{200}$. We also show the distributions of identified member galaxies for Abell~209, MACS~J0416, and MACS~J1206. 

The line-of-sight velocity dispersion profiles $\sigma_\text{los}(R)$ derived from each member galaxy sample serves as another important cross-validation of our membership identification scheme. As shown in the right panels of \fref{fig:Cluster_Membership_Identification}, our derived $\sigma_\text{los}(R)$ (cyan) are in excellent agreement with published projected stellar velocity dispersion profiles, available for Abell~383 \citep{Geller2014ApJ78352G}, Abell~209 \citep{Annunziatella2016AA585A160A}, and MACS~J1206 \citep{Biviano:2013eia}. The inference uncertainties in our derived profiles are model-agnostically estimated via bootstrap resampling with replacement from each full galaxy member catalog and quote the 10000-iteration $1\sigma$ dispersion (cyan error bars) \citep{Brown2010AJ13959B}.

\section{\Lenstool- vs. NFW-based CDM Truncation Radii Estimates}\label{app:Lenstool_vs_NFW}

\begin{figure}
	\includegraphics[width=0.99\linewidth]{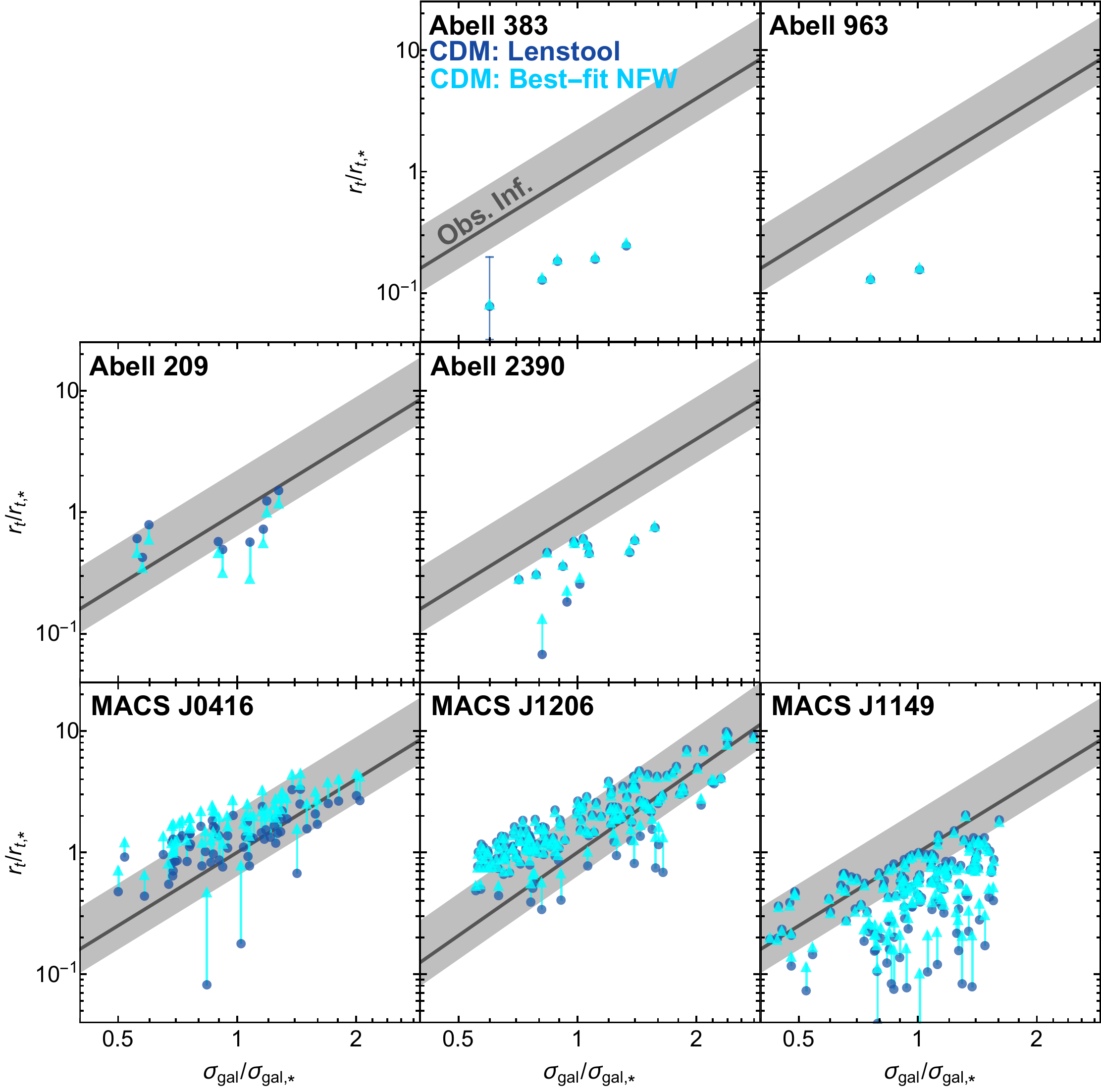}
    \caption{Tidal truncation radii of cluster subhalos derived from lensing-based observational inference (gray curves; gray-shading indicates conservative $5\sigma$) and CDM estimates from either $\Lenstool$-identified (blue circles, as in \fref{fig:CDM_SIDM}) or single best-fit NFW (cyan triangles) large-scale potentials for each cluster. We adopt $\tilde{\epsilon_i} = 0.47$; the exemplary error bar in the panel of Abell~383 shows the central 97\% range of subhalo-to-subhalo variance inferred from the TNG-Cluster analogs (\fref{fig:TNG_Cluster-3}). We observe an overall consistency between these two estimates; points showing the largest offsets are subhalos with the smallest projected distances to the respective cluster center and expectedly sensitive to the detailed large-scale potential modeling (i.e. multi-clump vs. single NFW cusp) in the strong lensing regime. In particular, the two clusters that show the most notable offsets\textemdash MACS~J0416 and J1149\textemdash have the largest projected separation between their two most massive clumps in our sample. Importantly, these two estimates are mutually compatible (within the subhalo-to-subhalo variance) and both statistically consistent with the observational inference, strengthening the robustness of our inferred SIDM constraints (\fref{fig:CDM_SIDM}).}
	\label{fig:Lenstool_Results_NFW}
\end{figure}

The large-scale background potential of massive clusters can and often comprises multiple large scale ``clumps'' \citep[e.g.][]{Natarajan:2002cw, Meneghetti2017MNRAS4723177M}, which differ from the strong+weak lensing best-fit \textit{single} NFW host potential per cluster commonly found in literature (e.g. see \tref{tab:Cluster_Properties} and \sref{sec:Cluster_Sample}). By construction, the former after shell-averaging and the latter converge at sufficiently large radii (\sref{ssec:rt_Estimates_Ana}), but this difference could potentially impact the robustness of our CDM/SIDM tidal radius estimates, especially for subhalos with small projected separation from the clump centers. In this Appendix, we demonstrate that such difference has minimal impact on our result (\fref{fig:CDM_SIDM}) by explicitly computing the CDM-predicted tidal radii by assuming cluster large potential provided by either the $\Lenstool$-optimized mass map (that permits multiple large-scale clumps) or best-fit single NFW potential. 

In the analysis presented \sref{sec:rt}, we preserve all the large-scale potentials identified by $\Lenstool$ for each cluster. The cluster's Center of Mass (CoM) is then computed as the arithmetic mean of projected coordinates of all large-scale potentials weighted by the respective enclosed mass within 500~kpc. These CoMs do differ but remain very close to the reported RA and DEC coordinates (\tref{tab:Cluster_Properties}) that serve as the CoMs of individual best-fit large-scale NFW potentials. Next for each cluster, we deproject each large-scale clump and construct a shell-averaged cluster density profile $\rho_\text{cluster}$ as in \eref{eqn:Obs_rt} centered on our computed CoM. Although accommodating for the possibility of multi-clump configuration, one approach does not account for ellipticity nor line-of-sight separation of these clumps. Aside from Abell~2218 already analyzed in \citet{Natarajan:2002cw}, we compare in \fref{fig:Lenstool_Results_NFW} the CDM-predicted tidal truncation radii under these two choices of large-scale cluster potential potentials. The overall high-level consistency between these two approaches further corroborates the robustness of our main result (\fref{fig:CDM_SIDM}).


\bibliographystyle{aasjournal}
\bibliography{MyBibTeX1} 

@ARTICLE{Fischer2024,
       author = {{Fischer}, Moritz S. and {Dolag}, Klaus and {Yu}, Hai-Bo},
        title = "{Numerical challenges for energy conservation in N-body simulations of collapsing self-interacting dark matter halos}",
      journal = {\aap},
         year = 2024,
        month = sep,
       volume = {689},
          eid = {A300},
        pages = {A300},
          doi = {10.1051/0004-6361/202449849},
archivePrefix = {arXiv},
       eprint = {2403.00739},
 primaryClass = {astro-ph.CO},
       adsurl = {https://ui.adsabs.harvard.edu/abs/2024A&A...689A.300F}
}

@article{lokas:hal-00005581,
  TITLE = {{Dark matter distribution in the Coma cluster from galaxy kinematics: breaking the mass-anisotropy degeneracy}},
  AUTHOR = {Lokas, Ewa L. and Mamon, Gary A.},
  URL = {https://hal.science/hal-00005581},
  NOTE = {13 pages, 10 figures, revised version accepted by MNRAS with discussion on substructure and equilibrium of elliptical galaxies added},
  JOURNAL = {{Monthly Notices of the Royal Astronomical Society}},
  PUBLISHER = {{Oxford University Press (OUP): Policy P - Oxford Open Option A}},
  VOLUME = {343},
  NUMBER = {2},
  PAGES = {401},
  YEAR = {2003},
  DOI = {10.1046/j.1365-8711.2003.06684.x},
  HAL_ID = {hal-00005581},
  HAL_VERSION = {v1},
}

@ARTICLE{Limousin2024,
       author = {{Limousin}, Marceau},
        title = "{Mass \& Light in Galaxy Clusters: Parametric Strong Lensing Approach}",
      journal = {arXiv e-prints},
         year = 2024,
        month = nov,
          eid = {arXiv:2411.03075},
        pages = {arXiv:2411.03075},
          doi = {10.48550/arXiv.2411.03075},
archivePrefix = {arXiv},
       eprint = {2411.03075},
 primaryClass = {astro-ph.CO},
       adsurl = {https://ui.adsabs.harvard.edu/abs/2024arXiv241103075L}
}

@software{2011ascl.soft02004K,
       author = {{Kneib}, Jean-Paul and {Bonnet}, Henri and {Golse}, Ghyslain and {Sand}, David and {Jullo}, Eric and {Marshall}, Phil},
        title = "{LENSTOOL: A Gravitational Lensing Software for Modeling Mass Distribution of Galaxies and Clusters (strong and weak regime)}",
 howpublished = {Astrophysics Source Code Library, record ascl:1102.004},
         year = 2011,
        month = feb,
          eid = {ascl:1102.004},
archivePrefix = {ascl},
       eprint = {1102.004},
       adsurl = {https://ui.adsabs.harvard.edu/abs/2011ascl.soft02004K}
}

@ARTICLE{Natarajan+2024,
       author = {{Natarajan}, P. and {Williams}, L.~L.~R. and {Brada{\v{c}}}, M. and {Grillo}, C. and {Ghosh}, A. and {Sharon}, K. and {Wagner}, J.},
        title = "{Strong Lensing by Galaxy Clusters}",
      journal = {\ssr},
         year = 2024,
        month = feb,
       volume = {220},
       number = {2},
          eid = {19},
        pages = {19},
          doi = {10.1007/s11214-024-01051-8},
archivePrefix = {arXiv},
       eprint = {2403.06245},
 primaryClass = {astro-ph.CO},
       adsurl = {https://ui.adsabs.harvard.edu/abs/2024SSRv..220...19N}
}

@ARTICLE{Bahcall2015,
       author = {{Bahcall}, Neta A.},
        title = "{Dark matter universe}",
      journal = {Proceedings of the National Academy of Science},
         year = 2015,
        month = oct,
       volume = {112},
       number = {40},
        pages = {12243-12245},
          doi = {10.1073/pnas.1516944112},
       adsurl = {https://ui.adsabs.harvard.edu/abs/2015PNAS..11212243B}
}

@article{Hall:2009bx,
	author = "Hall, Lawrence J. and Jedamzik, Karsten and March-Russell, John and West, Stephen M.",
	title = "{Freeze-In Production of FIMP Dark Matter}",
	eprint = "0911.1120",
	archivePrefix = "arXiv",
	primaryClass = "hep-ph",
	reportNumber = "OUTP-09-18-P, UCB-PTH-09-32",
	doi = "10.1007/JHEP03(2010)080",
	journal = "JHEP",
	volume = "03",
	pages = "080",
	year = "2010"
}

@article{Navarro:1996gj,
	author = "Navarro, Julio F. and Frenk, Carlos S. and White, Simon D.M.",
	title = "{A Universal density profile from hierarchical clustering}",
	eprint = "astro-ph/9611107",
	archivePrefix = "arXiv",
	doi = "10.1086/304888",
	journal = "Astrophys. J.",
	volume = "490",
	pages = "493--508",
	year = "1997"
}

@BOOK{Binney&Tremaine2008,
	author = {{Binney}, James and {Tremaine}, Scott},
	title = "{Galactic Dynamics: Second Edition}",
	year = 2008,
	adsurl = {https://ui.adsabs.harvard.edu/abs/2008gady.book.....B}
}

@article{Natarajan:2002cw,
	author = "Natarajan, Priyamvada and Loeb, Abraham and Kneib, Jean-Paul and Smail, Ian",
	title = "{Constraints on the collisional nature of the dark matter from gravitational lensing in the cluster a2218}",
	eprint = "astro-ph/0207045",
	archivePrefix = "arXiv",
	doi = "10.1086/345547",
	journal = "Astrophys. J. Lett.",
	volume = "580",
	pages = "L17--L20",
	year = "2002"
}

@article{Natarajan:1996yt,
	author = "Natarajan, Priyamvada and Kneib, Jean-Paul",
	title = "{Probing the dynamics of cluster-lenses}",
	eprint = "astro-ph/9602035",
	archivePrefix = "arXiv",
	doi = "10.1093/mnras/283.3.1031",
	journal = "Mon. Not. Roy. Astron. Soc.",
	volume = "283",
	pages = "1031",
	year = "1996"
}

@article{Furlanetto:2001tw,
	author = "Furlanetto, Steven and Loeb, Abraham",
	title = "{Constraining the collisional nature of the dark matter through observations of gravitational wakes}",
	eprint = "astro-ph/0107567",
	archivePrefix = "arXiv",
	doi = "10.1086/324693",
	journal = "Astrophys. J.",
	volume = "565",
	pages = "854",
	year = "2002"
}

@article{Tulin:2017ara,
	author = "Tulin, Sean and Yu, Hai-Bo",
	title = "{Dark Matter Self-interactions and Small Scale Structure}",
	eprint = "1705.02358",
	archivePrefix = "arXiv",
	primaryClass = "hep-ph",
	doi = "10.1016/j.physrep.2017.11.004",
	journal = "Phys. Rept.",
	volume = "730",
	pages = "1--57",
	year = "2018"
}

@ARTICLE{Nelson2019ComAC62N,
       author = {{Nelson}, Dylan and {Springel}, Volker and {Pillepich}, Annalisa and {Rodriguez-Gomez}, Vicente and {Torrey}, Paul and {Genel}, Shy and {Vogelsberger}, Mark and {Pakmor}, Ruediger and {Marinacci}, Federico and {Weinberger}, Rainer and {Kelley}, Luke and {Lovell}, Mark and {Diemer}, Benedikt and {Hernquist}, Lars},
        title = "{The IllustrisTNG simulations: public data release}",
      journal = {Computational Astrophysics and Cosmology},
         year = 2019,
        month = may,
       volume = {6},
       number = {1},
          eid = {2},
        pages = {2},
          doi = {10.1186/s40668-019-0028-x},
archivePrefix = {arXiv},
       eprint = {1812.05609},
 primaryClass = {astro-ph.GA},
       adsurl = {https://ui.adsabs.harvard.edu/abs/2019ComAC...6....2N}
}

@article{Randall:2008ppe,
	author = "Randall, Scott W. and Markevitch, Maxim and Clowe, Douglas and Gonzalez, Anthony H. and Bradac, Marusa",
	title = "{Constraints on the Self-Interaction Cross-Section of Dark Matter from Numerical Simulations of the Merging Galaxy Cluster 1E 0657-56}",
	eprint = "0704.0261",
	archivePrefix = "arXiv",
	primaryClass = "astro-ph",
	doi = "10.1086/587859",
	journal = "Astrophys. J.",
	volume = "679",
	pages = "1173--1180",
	year = "2008"
}

@article{Markevitch:2003at,
	author = "Markevitch, Maxim and Gonzalez, A. H. and Clowe, D. and Vikhlinin, A. and David, L. and Forman, W. and Jones, C. and Murray, S. and Tucker, W.",
	title = "{Direct constraints on the dark matter self-interaction cross-section from the merging galaxy cluster 1E0657-56}",
	eprint = "astro-ph/0309303",
	archivePrefix = "arXiv",
	doi = "10.1086/383178",
	journal = "Astrophys. J.",
	volume = "606",
	pages = "819--824",
	year = "2004"
}

@ARTICLE{Newman2013ApJ76524N,
	author = {{Newman}, Andrew B. and {Treu}, Tommaso and {Ellis}, Richard S. and {Sand}, David J. and {Nipoti}, Carlo and {Richard}, Johan and {Jullo}, Eric},
	title = "{The Density Profiles of Massive, Relaxed Galaxy Clusters. I. The Total Density Over Three Decades in Radius}",
	journal = "Astrophys. J.",
	year = 2013,
	month = mar,
	volume = {765},
	number = {1},
	eid = {24},
	pages = {24},
	doi = {10.1088/0004-637X/765/1/24},
	archivePrefix = {arXiv},
	eprint = {1209.1391},
	primaryClass = {astro-ph.CO},
	adsurl = {https://ui.adsabs.harvard.edu/abs/2013ApJ...765...24N}
}

@ARTICLE{LeBorgne1992AAS9587L,
	author = {{Le Borgne}, J.~F. and {Pello}, R. and {Sanahuja}, B.},
	title = "{Photometric and spectroscopic survey of the cluster of galaxies Abell 2218.}",
	journal = "Astron. Astrophys. J. Suppl.",
	year = 1992,
	month = oct,
	volume = {95},
	pages = {87-107},
	adsurl = {https://ui.adsabs.harvard.edu/abs/1992A&AS...95...87L}
}

@ARTICLE{Richard2010MNRAS404325R,
	author = {{Richard}, Johan and {Smith}, Graham P. and {Kneib}, Jean-Paul and {Ellis}, Richard S. and {Sanderson}, A.~J.~R. and {Pei}, L. and {Targett}, T.~A. and {Sand}, D.~J. and {Swinbank}, A.~M. and {Dannerbauer}, H. and {Mazzotta}, P. and {Limousin}, M. and {Egami}, E. and {Jullo}, E. and {Hamilton-Morris}, V. and {Moran}, S.~M.},
	title = "{LoCuSS: first results from strong-lensing analysis of 20 massive galaxy clusters at z = 0.2}",
	journal = "Mon. Not. Roy. Astron. Soc.",
	year = 2010,
	month = may,
	volume = {404},
	number = {1},
	pages = {325-349},
	doi = {10.1111/j.1365-2966.2009.16274.x},
	archivePrefix = {arXiv},
	eprint = {0911.3302},
	primaryClass = {astro-ph.CO},
	adsurl = {https://ui.adsabs.harvard.edu/abs/2010MNRAS.404..325R}
}

@article{Cerini:2022akj,
	author = "Cerini, Giulia and Cappelluti, Nico and Natarajan, Priyamvada",
	title = "{New Metrics to Probe the Dynamical State of Galaxy Clusters}",
	eprint = "2209.06831",
	archivePrefix = "arXiv",
	primaryClass = "astro-ph.CO",
	doi = "10.3847/1538-4357/acbccb",
	journal = "Astrophys. J.",
	volume = "945",
	number = "2",
	pages = "152",
	year = "2023"
}

@article{Harvey:2015hha,
	author = "Harvey, David and Massey, Richard and Kitching, Thomas and Taylor, Andy and Tittley, Eric",
	title = "{The non-gravitational interactions of dark matter in colliding galaxy clusters}",
	eprint = "1503.07675",
	archivePrefix = "arXiv",
	primaryClass = "astro-ph.CO",
	doi = "10.1126/science.1261381",
	journal = "Science",
	volume = "347",
	pages = "1462--1465",
	year = "2015"
}

@ARTICLE{Postman2012ApJS19925P,
	author = {{Postman}, Marc and {Coe}, Dan and {Ben{\'\i}tez}, Narciso and {Bradley}, Larry and {Broadhurst}, Tom and {Donahue}, Megan and {Ford}, Holland and {Graur}, Or and {Graves}, Genevieve and {Jouvel}, Stephanie and {Koekemoer}, Anton and {Lemze}, Doron and {Medezinski}, Elinor and {Molino}, Alberto and {Moustakas}, Leonidas and {Ogaz}, Sara and {Riess}, Adam and {Rodney}, Steve and {Rosati}, Piero and {Umetsu}, Keiichi and {Zheng}, Wei and {Zitrin}, Adi and {Bartelmann}, Matthias and {Bouwens}, Rychard and {Czakon}, Nicole and {Golwala}, Sunil and {Host}, Ole and {Infante}, Leopoldo and {Jha}, Saurabh and {Jimenez-Teja}, Yolanda and {Kelson}, Daniel and {Lahav}, Ofer and {Lazkoz}, Ruth and {Maoz}, Dani and {McCully}, Curtis and {Melchior}, Peter and {Meneghetti}, Massimo and {Merten}, Julian and {Moustakas}, John and {Nonino}, Mario and {Patel}, Brandon and {Reg{\"o}s}, Enik{\"o} and {Sayers}, Jack and {Seitz}, Stella and {Van der Wel}, Arjen},
	title = "{The Cluster Lensing and Supernova Survey with Hubble: An Overview}",
	journal = "Astrophys. J. Suppl.",
	year = 2012,
	month = apr,
	volume = {199},
	number = {2},
	eid = {25},
	pages = {25},
	doi = {10.1088/0067-0049/199/2/25},
	archivePrefix = {arXiv},
	eprint = {1106.3328},
	primaryClass = {astro-ph.CO},
	adsurl = {https://ui.adsabs.harvard.edu/abs/2012ApJS..199...25P}
}

@article{Girardi:2015aga,
	author = "Girardi, M. and others",
	title = "{CLASH-VLT: Substructure in the galaxy cluster MACS J1206.2-0847 from kinematics of galaxy populations}",
	eprint = "1503.05607",
	archivePrefix = "arXiv",
	primaryClass = "astro-ph.CO",
	doi = "10.1051/0004-6361/201425599",
	journal = "Astron. Astrophys.",
	volume = "579",
	pages = "A4",
	year = "2015"
}

@ARTICLE{Umetsu2012ApJ75556U,
	author = {{Umetsu}, Keiichi and {Medezinski}, Elinor and {Nonino}, Mario and {Merten}, Julian and {Zitrin}, Adi and {Molino}, Alberto and {Grillo}, Claudio and {Carrasco}, Mauricio and {Donahue}, Megan and {Mahdavi}, Andisheh and {Coe}, Dan and {Postman}, Marc and {Koekemoer}, Anton and {Czakon}, Nicole and {Sayers}, Jack and {Mroczkowski}, Tony and {Golwala}, Sunil and {Koch}, Patrick M. and {Lin}, Kai-Yang and {Molnar}, Sandor M. and {Rosati}, Piero and {Balestra}, Italo and {Mercurio}, Amata and {Scodeggio}, Marco and {Biviano}, Andrea and {Anguita}, Timo and {Infante}, Leopoldo and {Seidel}, Gregor and {Sendra}, Irene and {Jouvel}, Stephanie and {Host}, Ole and {Lemze}, Doron and {Broadhurst}, Tom and {Meneghetti}, Massimo and {Moustakas}, Leonidas and {Bartelmann}, Matthias and {Ben{\'\i}tez}, Narciso and {Bouwens}, Rychard and {Bradley}, Larry and {Ford}, Holland and {Jim{\'e}nez-Teja}, Yolanda and {Kelson}, Daniel and {Lahav}, Ofer and {Melchior}, Peter and {Moustakas}, John and {Ogaz}, Sara and {Seitz}, Stella and {Zheng}, Wei},
	title = "{CLASH: Mass Distribution in and around MACS J1206.2-0847 from a Full Cluster Lensing Analysis}",
	journal = "Astrophys. J.",
	year = 2012,
	month = aug,
	volume = {755},
	number = {1},
	eid = {56},
	pages = {56},
	doi = {10.1088/0004-637X/755/1/56},
	archivePrefix = {arXiv},
	eprint = {1204.3630},
	primaryClass = {astro-ph.CO},
	adsurl = {https://ui.adsabs.harvard.edu/abs/2012ApJ...755...56U}
}

@article{Biviano:2013eia,
	author = "Biviano, A. and others",
	title = "{CLASH-VLT: The mass, velocity-anisotropy, and pseudo-phase-space density profiles of the z=0.44 galaxy cluster MACS 1206.2-0847}",
	eprint = "1307.5867",
	archivePrefix = "arXiv",
	primaryClass = "astro-ph.CO",
	doi = "10.1051/0004-6361/201321955",
	journal = "Astron. Astrophys.",
	volume = "558",
	pages = "A1",
	year = "2013"
}

@ARTICLE{Ueda2020ApJ892100U,
	title = "{Gas Density Perturbations in the Cool Cores of CLASH Galaxy Clusters}",
	journal = "Astrophys. J.",
	year = 2020,
	month = apr,
	volume = {892},
	number = {2},
	eid = {100},
	pages = {100},
	doi = {10.3847/1538-4357/ab7bdc},
	archivePrefix = {arXiv},
	eprint = {1912.07300},
	primaryClass = {astro-ph.CO},
	adsurl = {https://ui.adsabs.harvard.edu/abs/2020ApJ...892..100U}
}

@ARTICLE{Biviano2023ApJ958148B,
	author = {{Biviano}, Andrea and {Pizzuti}, Lorenzo and {Mercurio}, Amata and {Sartoris}, Barbara and {Rosati}, Piero and {Ettori}, Stefano and {Girardi}, Marisa and {Grillo}, Claudio and {Caminha}, Gabriel B. and {Nonino}, Mario},
	title = "{CLASH-VLT: The Inner Slope of the MACS J1206.2-0847 Dark Matter Density Profile}",
	journal = {\apj},
	year = 2023,
	month = dec,
	volume = {958},
	number = {2},
	eid = {148},
	pages = {148},
	doi = {10.3847/1538-4357/acf832},
	archivePrefix = {arXiv},
	eprint = {2307.06804},
	primaryClass = {astro-ph.CO},
	adsurl = {https://ui.adsabs.harvard.edu/abs/2023ApJ...958..148B}
}

@ARTICLE{Smith2005MNRAS359417S,
	author = {{Smith}, Graham P. and {Kneib}, Jean-Paul and {Smail}, Ian and {Mazzotta}, Pasquale and {Ebeling}, Harald and {Czoske}, Oliver},
	title = "{A Hubble Space Telescope lensing survey of X-ray luminous galaxy clusters - IV. Mass, structure and thermodynamics of cluster cores at z= 0.2}",
	journal = "Mon. Not. Roy. Astron. Soc.",
	year = 2005,
	month = may,
	volume = {359},
	number = {2},
	pages = {417-446},
	doi = {10.1111/j.1365-2966.2005.08911.x},
	archivePrefix = {arXiv},
	eprint = {astro-ph/0403588},
	primaryClass = {astro-ph},
	adsurl = {https://ui.adsabs.harvard.edu/abs/2005MNRAS.359..417S}
}

@ARTICLE{Gilmour2009MNRAS3921509G,
	author = {{Gilmour}, R. and {Best}, P. and {Almaini}, O.},
	title = "{The distribution of active galactic nuclei in a large sample of galaxy clusters}",
	journal = "Mon. Not. Roy. Astron. Soc.",
	year = 2009,
	month = feb,
	volume = {392},
	number = {4},
	pages = {1509-1531},
	doi = {10.1111/j.1365-2966.2008.14161.x},
	archivePrefix = {arXiv},
	eprint = {0901.2810},
	primaryClass = {astro-ph.GA},
	adsurl = {https://ui.adsabs.harvard.edu/abs/2009MNRAS.392.1509G}
}

@ARTICLE{Annunziatella2016AA585A160A,
	author = {{Annunziatella}, M. and {Mercurio}, A. and {Biviano}, A. and {Girardi}, M. and {Nonino}, M. and {Balestra}, I. and {Rosati}, P. and {Bartosch Caminha}, G. and {Brescia}, M. and {Gobat}, R. and {Grillo}, C. and {Lombardi}, M. and {Sartoris}, B. and {De Lucia}, G. and {Demarco}, R. and {Frye}, B. and {Fritz}, A. and {Moustakas}, J. and {Scodeggio}, M. and {Kuchner}, U. and {Maier}, C. and {Ziegler}, B.},
	title = "{CLASH-VLT: Environment-driven evolution of galaxies in the z = 0.209 cluster Abell 209}",
	journal = {Astron. Astrophys.},
	year = 2016,
	month = jan,
	volume = {585},
	eid = {A160},
	pages = {A160},
	doi = {10.1051/0004-6361/201527399},
	archivePrefix = {arXiv},
	eprint = {1510.05659},
	primaryClass = {astro-ph.GA},
	adsurl = {https://ui.adsabs.harvard.edu/abs/2016A&A...585A.160A}
}

@ARTICLE{Balestra2016ApJS22433B,
	author = {{Balestra}, I. and {Mercurio}, A. and {Sartoris}, B. and {Girardi}, M. and {Grillo}, C. and {Nonino}, M. and {Rosati}, P. and {Biviano}, A. and {Ettori}, S. and {Forman}, W. and {Jones}, C. and {Koekemoer}, A. and {Medezinski}, E. and {Merten}, J. and {Ogrean}, G.~A. and {Tozzi}, P. and {Umetsu}, K. and {Vanzella}, E. and {van Weeren}, R.~J. and {Zitrin}, A. and {Annunziatella}, M. and {Caminha}, G.~B. and {Broadhurst}, T. and {Coe}, D. and {Donahue}, M. and {Fritz}, A. and {Frye}, B. and {Kelson}, D. and {Lombardi}, M. and {Maier}, C. and {Meneghetti}, M. and {Monna}, A. and {Postman}, M. and {Scodeggio}, M. and {Seitz}, S. and {Ziegler}, B.},
	title = "{CLASH-VLT: Dissecting the Frontier Fields Galaxy Cluster MACS J0416.1-2403 with {\ensuremath{\sim}}800 Spectra of Member Galaxies}",
	journal = "Astrophys. J. Suppl.",
	year = 2016,
	month = jun,
	volume = {224},
	number = {2},
	eid = {33},
	pages = {33},
	doi = {10.3847/0067-0049/224/2/33},
	archivePrefix = {arXiv},
	eprint = {1511.02522},
	primaryClass = {astro-ph.CO},
	adsurl = {https://ui.adsabs.harvard.edu/abs/2016ApJS..224...33B}
}

@ARTICLE{Caminha2017AA600A90C,
	author = {{Caminha}, G.~B. and {Grillo}, C. and {Rosati}, P. and {Balestra}, I. and {Mercurio}, A. and {Vanzella}, E. and {Biviano}, A. and {Caputi}, K.~I. and {Delgado-Correal}, C. and {Karman}, W. and {Lombardi}, M. and {Meneghetti}, M. and {Sartoris}, B. and {Tozzi}, P.},
	title = "{A refined mass distribution of the cluster MACS J0416.1-2403 from a new large set of spectroscopic multiply lensed sources}",
	journal = {Astron. Astrophys.},
	year = 2017,
	month = apr,
	volume = {600},
	eid = {A90},
	pages = {A90},
	doi = {10.1051/0004-6361/201629297},
	archivePrefix = {arXiv},
	eprint = {1607.03462},
	primaryClass = {astro-ph.GA},
	adsurl = {https://ui.adsabs.harvard.edu/abs/2017A&A...600A..90C}
}

@ARTICLE{Grillo2016ApJ82278G,
	author = {{Grillo}, C. and {Karman}, W. and {Suyu}, S.~H. and {Rosati}, P. and {Balestra}, I. and {Mercurio}, A. and {Lombardi}, M. and {Treu}, T. and {Caminha}, G.~B. and {Halkola}, A. and {Rodney}, S.~A. and {Gavazzi}, R. and {Caputi}, K.~I.},
	title = "{The Story of Supernova {\textquotedblleft}Refsdal{\textquotedblright} Told by Muse}",
	journal = "Astrophys. J.",
	year = 2016,
	month = may,
	volume = {822},
	number = {2},
	eid = {78},
	pages = {78},
	doi = {10.3847/0004-637X/822/2/78},
	archivePrefix = {arXiv},
	eprint = {1511.04093},
	primaryClass = {astro-ph.GA},
	adsurl = {https://ui.adsabs.harvard.edu/abs/2016ApJ...822...78G}
}

@ARTICLE{Bergamini2019AA631A130B,
	author = {{Bergamini}, P. and {Rosati}, P. and {Mercurio}, A. and {Grillo}, C. and {Caminha}, G.~B. and {Meneghetti}, M. and {Agnello}, A. and {Biviano}, A. and {Calura}, F. and {Giocoli}, C. and {Lombardi}, M. and {Rodighiero}, G. and {Vanzella}, E.},
	title = "{Enhanced cluster lensing models with measured galaxy kinematics}",
	journal = {Astron. Astrophys.},
	year = 2019,
	month = nov,
	volume = {631},
	eid = {A130},
	pages = {A130},
	doi = {10.1051/0004-6361/201935974},
	archivePrefix = {arXiv},
	eprint = {1905.13236},
	primaryClass = {astro-ph.GA},
	adsurl = {https://ui.adsabs.harvard.edu/abs/2019A&A...631A.130B}
}

@ARTICLE{Finney2018ApJ85958F,
	author = {{Finney}, Emily Quinn and {Brada{\v{c}}}, Maru{\v{s}}a and {Huang}, Kuang-Han and {Hoag}, Austin and {Morishita}, Takahiro and {Schrabback}, Tim and {Treu}, Tommaso and {Borello Schmidt}, Kasper and {Lemaux}, Brian C. and {Wang}, Xin and {Mason}, Charlotte},
	title = "{Mass Modeling of Frontier Fields Cluster MACS J1149.5+2223 Using Strong and Weak Lensing}",
	journal = "Astrophys. J.",
	year = 2018,
	month = may,
	volume = {859},
	number = {1},
	eid = {58},
	pages = {58},
	doi = {10.3847/1538-4357/aabf97},
	adsurl = {https://ui.adsabs.harvard.edu/abs/2018ApJ...859...58F}
}

@ARTICLE{Merten2015ApJ8064M,
	author = {{Merten}, J. and {Meneghetti}, M. and {Postman}, M. and {Umetsu}, K. and {Zitrin}, A. and {Medezinski}, E. and {Nonino}, M. and {Koekemoer}, A. and {Melchior}, P. and {Gruen}, D. and {Moustakas}, L.~A. and {Bartelmann}, M. and {Host}, O. and {Donahue}, M. and {Coe}, D. and {Molino}, A. and {Jouvel}, S. and {Monna}, A. and {Seitz}, S. and {Czakon}, N. and {Lemze}, D. and {Sayers}, J. and {Balestra}, I. and {Rosati}, P. and {Ben{\'\i}tez}, N. and {Biviano}, A. and {Bouwens}, R. and {Bradley}, L. and {Broadhurst}, T. and {Carrasco}, M. and {Ford}, H. and {Grillo}, C. and {Infante}, L. and {Kelson}, D. and {Lahav}, O. and {Massey}, R. and {Moustakas}, J. and {Rasia}, E. and {Rhodes}, J. and {Vega}, J. and {Zheng}, W.},
	title = "{CLASH: The Concentration-Mass Relation of Galaxy Clusters}",
	journal = "Astrophys. J.",
	year = 2015,
	month = jun,
	volume = {806},
	number = {1},
	eid = {4},
	pages = {4},
	doi = {10.1088/0004-637X/806/1/4},
	archivePrefix = {arXiv},
	eprint = {1404.1376},
	primaryClass = {astro-ph.CO},
	adsurl = {https://ui.adsabs.harvard.edu/abs/2015ApJ...806....4M}
}

@ARTICLE{Biviano2021A&A650A105B,
	author = {{Biviano}, A. and {van der Burg}, R.~F.~J. and {Balogh}, M.~L. and {Munari}, E. and {Cooper}, M.~C. and {De Lucia}, G. and {Demarco}, R. and {Jablonka}, P. and {Muzzin}, A. and {Nantais}, J. and {Old}, L.~J. and {Rudnick}, G. and {Vulcani}, B. and {Wilson}, G. and {Yee}, H.~K.~C. and {Zaritsky}, D. and {Cerulo}, P. and {Chan}, J. and {Finoguenov}, A. and {Gilbank}, D. and {Lidman}, C. and {Pintos-Castro}, I. and {Shipley}, H.},
	title = "{The GOGREEN survey: Internal dynamics of clusters of galaxies at redshift 0.9-1.4}",
	journal = {\aap},
	year = 2021,
	month = jun,
	volume = {650},
	eid = {A105},
	pages = {A105},
	doi = {10.1051/0004-6361/202140564},
	archivePrefix = {arXiv},
	eprint = {2104.01183},
	primaryClass = {astro-ph.CO},
	adsurl = {https://ui.adsabs.harvard.edu/abs/2021A&A...650A.105B}
}

@ARTICLE{Olave-Rojas2023MNRAS5194171O,
	author = {{Olave-Rojas}, D.~E. and {Cerulo}, P. and {Araya-Araya}, P. and {Olave-Rojas}, D.~A.},
	title = "{CALSAGOS: Clustering algorithms applied to galaxies in overdense systems}",
	journal = {\mnras},
	year = 2023,
	month = mar,
	volume = {519},
	number = {3},
	pages = {4171-4182},
	doi = {10.1093/mnras/stac3762},
	archivePrefix = {arXiv},
	eprint = {2212.09451},
	primaryClass = {astro-ph.GA},
	adsurl = {https://ui.adsabs.harvard.edu/abs/2023MNRAS.519.4171O}
}

@ARTICLE{Geller2014ApJ78352G,
	author = {{Geller}, Margaret J. and {Hwang}, Ho Seong and {Diaferio}, Antonaldo and {Kurtz}, Michael J. and {Coe}, Dan and {Rines}, Kenneth J.},
	title = "{A Redshift Survey of the Strong-lensing Cluster Abell 383}",
	journal = {\apj},
	year = 2014,
	month = mar,
	volume = {783},
	number = {1},
	eid = {52},
	pages = {52},
	doi = {10.1088/0004-637X/783/1/52},
	archivePrefix = {arXiv},
	eprint = {1401.1440},
	primaryClass = {astro-ph.CO},
	adsurl = {https://ui.adsabs.harvard.edu/abs/2014ApJ...783...52G}
}

@ARTICLE{Rines2016ApJ81963R,
	author = {{Rines}, Kenneth J. and {Geller}, Margaret J. and {Diaferio}, Antonaldo and {Hwang}, Ho Seong},
	title = "{HeCS-SZ: The Hectospec Survey of Sunyaev-Zeldovich-selected Clusters}",
	journal = {\apj},
	year = 2016,
	month = mar,
	volume = {819},
	number = {1},
	eid = {63},
	pages = {63},
	doi = {10.3847/0004-637X/819/1/63},
	archivePrefix = {arXiv},
	eprint = {1507.08289},
	primaryClass = {astro-ph.CO},
	adsurl = {https://ui.adsabs.harvard.edu/abs/2016ApJ...819...63R}
}

@ARTICLE{Koulouridis2021A&A652A12K,
	author = {{Koulouridis}, E. and {Clerc}, N. and {Sadibekova}, T. and {Chira}, M. and {Drigga}, E. and {Faccioli}, L. and {Le F{\`e}vre}, J.~P. and {Garrel}, C. and {Gaynullina}, E. and {Gkini}, A. and {Kosiba}, M. and {Pacaud}, F. and {Pierre}, M. and {Ridl}, J. and {Tazhenova}, K. and {Adami}, C. and {Altieri}, B. and {Baguley}, J. -C. and {Cabanac}, R. and {Cucchetti}, E. and {Khalikova}, A. and {Lieu}, M. and {Melin}, J. -B. and {Molham}, M. and {Ramos-Ceja}, M.~E. and {Soucail}, G. and {Takey}, A. and {Valtchanov}, I.},
	title = "{The X-CLASS survey: A catalogue of 1646 X-ray-selected galaxy clusters up to z {\ensuremath{\sim}} 1.5}",
	journal = {\aap},
	year = 2021,
	month = aug,
	volume = {652},
	eid = {A12},
	pages = {A12},
	doi = {10.1051/0004-6361/202140566},
	archivePrefix = {arXiv},
	eprint = {2104.06617},
	primaryClass = {astro-ph.CO},
	adsurl = {https://ui.adsabs.harvard.edu/abs/2021A&A...652A..12K}
}

@ARTICLE{Umetsu2016ApJ821116U,
	author = {{Umetsu}, Keiichi and {Zitrin}, Adi and {Gruen}, Daniel and {Merten}, Julian and {Donahue}, Megan and {Postman}, Marc},
	title = "{CLASH: Joint Analysis of Strong-lensing, Weak-lensing Shear, and Magnification Data for 20 Galaxy Clusters}",
	journal = {\apj},
	year = 2016,
	month = apr,
	volume = {821},
	number = {2},
	eid = {116},
	pages = {116},
	doi = {10.3847/0004-637X/821/2/116},
	archivePrefix = {arXiv},
	eprint = {1507.04385},
	primaryClass = {astro-ph.CO},
	adsurl = {https://ui.adsabs.harvard.edu/abs/2016ApJ...821..116U}
}

@ARTICLE{Lotz2017ApJ83797L,
	author = {{Lotz}, J.~M. and {Koekemoer}, A. and {Coe}, D. and {Grogin}, N. and {Capak}, P. and {Mack}, J. and {Anderson}, J. and {Avila}, R. and {Barker}, E.~A. and {Borncamp}, D. and {Brammer}, G. and {Durbin}, M. and {Gunning}, H. and {Hilbert}, B. and {Jenkner}, H. and {Khandrika}, H. and {Levay}, Z. and {Lucas}, R.~A. and {MacKenty}, J. and {Ogaz}, S. and {Porterfield}, B. and {Reid}, N. and {Robberto}, M. and {Royle}, P. and {Smith}, L.~J. and {Storrie-Lombardi}, L.~J. and {Sunnquist}, B. and {Surace}, J. and {Taylor}, D.~C. and {Williams}, R. and {Bullock}, J. and {Dickinson}, M. and {Finkelstein}, S. and {Natarajan}, P. and {Richard}, J. and {Robertson}, B. and {Tumlinson}, J. and {Zitrin}, A. and {Flanagan}, K. and {Sembach}, K. and {Soifer}, B.~T. and {Mountain}, M.},
	title = "{The Frontier Fields: Survey Design and Initial Results}",
	journal = {\apj},
	year = 2017,
	month = mar,
	volume = {837},
	number = {1},
	eid = {97},
	pages = {97},
	doi = {10.3847/1538-4357/837/1/97},
	archivePrefix = {arXiv},
	eprint = {1605.06567},
	primaryClass = {astro-ph.GA},
	adsurl = {https://ui.adsabs.harvard.edu/abs/2017ApJ...837...97L}
}

@ARTICLE{Natarajan1997MNRAS287833N,
	author = {{Natarajan}, Priyamvada and {Kneib}, Jean-Paul},
	title = "{Lensing by galaxy haloes in clusters of galaxies}",
	journal = {\mnras},
	year = 1997,
	month = jun,
	volume = {287},
	number = {4},
	pages = {833-847},
	doi = {10.1093/mnras/287.4.833},
	archivePrefix = {arXiv},
	eprint = {astro-ph/9609008},
	primaryClass = {astro-ph},
	adsurl = {https://ui.adsabs.harvard.edu/abs/1997MNRAS.287..833N}
}

@ARTICLE{Kneib2011A&ARv1947K,
	author = {{Kneib}, Jean-Paul and {Natarajan}, Priyamvada},
	title = "{Cluster lenses}",
	journal = {\aapr},
	year = 2011,
	month = nov,
	volume = {19},
	eid = {47},
	pages = {47},
	doi = {10.1007/s00159-011-0047-3},
	archivePrefix = {arXiv},
	eprint = {1202.0185},
	primaryClass = {astro-ph.CO},
	adsurl = {https://ui.adsabs.harvard.edu/abs/2011A&ARv..19...47K}
}

@ARTICLE{Brown2010AJ13959B,
	author = {{Brown}, Warren R. and {Geller}, Margaret J. and {Kenyon}, Scott J. and {Diaferio}, Antonaldo},
	title = "{Velocity Dispersion Profile of the Milky Way Halo}",
	journal = {\aj},
	year = 2010,
	month = jan,
	volume = {139},
	number = {1},
	pages = {59-67},
	doi = {10.1088/0004-6256/139/1/59},
	archivePrefix = {arXiv},
	eprint = {0910.2242},
	primaryClass = {astro-ph.GA},
	adsurl = {https://ui.adsabs.harvard.edu/abs/2010AJ....139...59B}
}

@ARTICLE{Schechter1976ApJ203297S,
	author = {{Schechter}, P.},
	title = "{An analytic expression for the luminosity function for galaxies.}",
	journal = {\apj},
	year = 1976,
	month = jan,
	volume = {203},
	pages = {297-306},
	doi = {10.1086/154079},
	adsurl = {https://ui.adsabs.harvard.edu/abs/1976ApJ...203..297S}
}

@ARTICLE{Banerjee2020JCAP02024B,
	author = {{Banerjee}, Arka and {Adhikari}, Susmita and {Dalal}, Neal and {More}, Surhud and {Kravtsov}, Andrey},
	title = "{Signatures of self-interacting dark matter on cluster density profile and subhalo distributions}",
	journal = {\jcap},
	year = 2020,
	month = feb,
	volume = {2020},
	number = {2},
	eid = {024},
	pages = {024},
	doi = {10.1088/1475-7516/2020/02/024},
	archivePrefix = {arXiv},
	eprint = {1906.12026},
	primaryClass = {astro-ph.CO},
	adsurl = {https://ui.adsabs.harvard.edu/abs/2020JCAP...02..024B}
}

@ARTICLE{Eliasdottir2007arXiv07105636E,
	author = {{El{\'\i}asd{\'o}ttir}, {\'A}rd{\'\i}s and {Limousin}, Marceau and {Richard}, Johan and {Hjorth}, Jens and {Kneib}, Jean-Paul and {Natarajan}, Priya and {Pedersen}, Kristian and {Jullo}, Eric and {Paraficz}, Danuta},
	title = "{Where is the matter in the Merging Cluster Abell 2218?}",
	journal = {arXiv e-prints},
	year = 2007,
	month = oct,
	eid = {arXiv:0710.5636},
	pages = {arXiv:0710.5636},
	doi = {10.48550/arXiv.0710.5636},
	archivePrefix = {arXiv},
	eprint = {0710.5636},
	primaryClass = {astro-ph},
	adsurl = {https://ui.adsabs.harvard.edu/abs/2007arXiv0710.5636E}
}

@ARTICLE{Dutra2025ApJ97838D,
	author = {{Dutra}, Isaque and {Natarajan}, Priyamvada and {Gilman}, Daniel},
	title = "{Self-interacting Dark Matter, Core Collapse, and the Galaxy{\textendash}Galaxy Strong-lensing Discrepancy}",
	journal = {\apj},
	year = 2025,
	month = jan,
	volume = {978},
	number = {1},
	eid = {38},
	pages = {38},
	doi = {10.3847/1538-4357/ad9b09},
	archivePrefix = {arXiv},
	eprint = {2406.17024},
	primaryClass = {astro-ph.CO},
	adsurl = {https://ui.adsabs.harvard.edu/abs/2025ApJ...978...38D}
}

@ARTICLE{Xu2022AA658A59X,
	author = {{Xu}, Weiwei and {Ramos-Ceja}, Miriam E. and {Pacaud}, Florian and {Reiprich}, Thomas H. and {Erben}, Thomas},
	title = "{Catalog of X-ray-selected extended galaxy clusters from the ROSAT All-Sky Survey (RXGCC)}",
	journal = {Astron. Astrophys.},
	year = 2022,
	month = feb,
	volume = {658},
	eid = {A59},
	pages = {A59},
	doi = {10.1051/0004-6361/202140908},
	archivePrefix = {arXiv},
	eprint = {2110.14886},
	primaryClass = {astro-ph.CO},
	adsurl = {https://ui.adsabs.harvard.edu/abs/2022A&A...658A..59X}
}

@ARTICLE{Mahdavi2001ApJ554L129M,
	author = {{Mahdavi}, Andisheh and {Geller}, Margaret J.},
	title = "{The L$_{X}$-{\ensuremath{\sigma}} Relation for Galaxies and Clusters of Galaxies}",
	journal = {\apjl},
	year = 2001,
	month = jun,
	volume = {554},
	number = {2},
	pages = {L129-L132},
	doi = {10.1086/321710},
	archivePrefix = {arXiv},
	eprint = {astro-ph/0105315},
	primaryClass = {astro-ph},
	adsurl = {https://ui.adsabs.harvard.edu/abs/2001ApJ...554L.129M}
}

@ARTICLE{Cannon1999MNRAS3029C,
	author = {{Cannon}, D.~B. and {Ponman}, T.~J. and {Hobbs}, I.~S.},
	title = "{The mass and dynamical state of Abell 2218}",
	journal = {\mnras},
	year = 1999,
	month = jan,
	volume = {302},
	number = {1},
	pages = {9-21},
	doi = {10.1046/j.1365-8711.1999.01886.x},
	archivePrefix = {arXiv},
	eprint = {astro-ph/9806230},
	primaryClass = {astro-ph},
	adsurl = {https://ui.adsabs.harvard.edu/abs/1999MNRAS.302....9C}
}

@ARTICLE{Wenger2000AAS1439W,
	author = {{Wenger}, M. and {Ochsenbein}, F. and {Egret}, D. and {Dubois}, P. and {Bonnarel}, F. and {Borde}, S. and {Genova}, F. and {Jasniewicz}, G. and {Lalo{\"e}}, S. and {Lesteven}, S. and {Monier}, R.},
	title = "{The SIMBAD astronomical database. The CDS reference database for astronomical objects}",
	journal = {\aaps},
	year = 2000,
	month = apr,
	volume = {143},
	pages = {9-22},
	doi = {10.1051/aas:2000332},
	archivePrefix = {arXiv},
	eprint = {astro-ph/0002110},
	primaryClass = {astro-ph},
	adsurl = {https://ui.adsabs.harvard.edu/abs/2000A&AS..143....9W}
}

@INPROCEEDINGS{Marty2003SPIE4851208M,
	author = {{Marty}, Philippe B. and {Kneib}, Jean-Paul and {Sadat}, Rachida and {Ebeling}, Harald and {Smail}, Ian},
	title = "{Data analysis method for XMM-Newton observations of extended sources and application to bright massive clusters of galaxies at z=0.2}",
	booktitle = {X-Ray and Gamma-Ray Telescopes and Instruments for Astronomy.},
	year = 2003,
	editor = {{Truemper}, Joachim E. and {Tananbaum}, Harvey D.},
	series = {Society of Photo-Optical Instrumentation Engineers (SPIE) Conference Series},
	volume = {4851},
	month = mar,
	pages = {208-222},
	doi = {10.1117/12.461330},
	archivePrefix = {arXiv},
	eprint = {astro-ph/0209270},
	primaryClass = {astro-ph},
	adsurl = {https://ui.adsabs.harvard.edu/abs/2003SPIE.4851..208M}
}

@ARTICLE{Natarajan2017MNRAS4681962N,
	author = {{Natarajan}, Priyamvada and {Chadayammuri}, Urmila and {Jauzac}, Mathilde and {Richard}, Johan and {Kneib}, Jean-Paul and {Ebeling}, Harald and {Jiang}, Fangzhou and {van den Bosch}, Frank and {Limousin}, Marceau and {Jullo}, Eric and {Atek}, Hakim and {Pillepich}, Annalisa and {Popa}, Cristina and {Marinacci}, Federico and {Hernquist}, Lars and {Meneghetti}, Massimo and {Vogelsberger}, Mark},
	title = "{Mapping substructure in the HST Frontier Fields cluster lenses and in cosmological simulations}",
	journal = {\mnras},
	year = 2017,
	month = jun,
	volume = {468},
	number = {2},
	pages = {1962-1980},
	doi = {10.1093/mnras/stw3385},
	archivePrefix = {arXiv},
	eprint = {1702.04348},
	primaryClass = {astro-ph.GA},
	adsurl = {https://ui.adsabs.harvard.edu/abs/2017MNRAS.468.1962N}
}

@ARTICLE{Caminha2017AA607A93C,
	author = {{Caminha}, G.~B. and {Grillo}, C. and {Rosati}, P. and {Meneghetti}, M. and {Mercurio}, A. and {Ettori}, S. and {Balestra}, I. and {Biviano}, A. and {Umetsu}, K. and {Vanzella}, E. and {Annunziatella}, M. and {Bonamigo}, M. and {Delgado-Correal}, C. and {Girardi}, M. and {Lombardi}, M. and {Nonino}, M. and {Sartoris}, B. and {Tozzi}, P. and {Bartelmann}, M. and {Bradley}, L. and {Caputi}, K.~I. and {Coe}, D. and {Ford}, H. and {Fritz}, A. and {Gobat}, R. and {Postman}, M. and {Seitz}, S. and {Zitrin}, A.},
	title = "{Mass distribution in the core of MACS J1206. Robust modeling from an exceptionally large sample of central multiple images}",
	journal = {\aap},
	year = 2017,
	month = nov,
	volume = {607},
	eid = {A93},
	pages = {A93},
	doi = {10.1051/0004-6361/201731498},
	archivePrefix = {arXiv},
	eprint = {1707.00690},
	primaryClass = {astro-ph.GA},
	adsurl = {https://ui.adsabs.harvard.edu/abs/2017A&A...607A..93C}
}

@ARTICLE{Kneib1996ApJ471643K,
	author = {{Kneib}, J. -P. and {Ellis}, R.~S. and {Smail}, I. and {Couch}, W.~J. and {Sharples}, R.~M.},
	title = "{Hubble Space Telescope Observations of the Lensing Cluster Abell 2218}",
	journal = {\apj},
	year = 1996,
	month = nov,
	volume = {471},
	pages = {643},
	doi = {10.1086/177995},
	archivePrefix = {arXiv},
	eprint = {astro-ph/9511015},
	primaryClass = {astro-ph},
	adsurl = {https://ui.adsabs.harvard.edu/abs/1996ApJ...471..643K}
}

@article{Meneghetti:2020yif,
    author = "Meneghetti, Massimo and others",
    title = "{An excess of small-scale gravitational lenses observed in galaxy clusters}",
    eprint = "2009.04471",
    archivePrefix = "arXiv",
    primaryClass = "astro-ph.GA",
    doi = "10.1126/science.aax5164",
    journal = "Science",
    volume = "369",
    number = "6509",
    pages = "1347--1351",
    year = "2020"
}

@article{Meneghetti:2022apr,
    author = "Meneghetti, Massimo and others",
    title = "{The probability of galaxy-galaxy strong lensing events in hydrodynamical simulations of galaxy clusters}",
    eprint = "2204.09065",
    archivePrefix = "arXiv",
    primaryClass = "astro-ph.CO",
    doi = "10.1051/0004-6361/202243779",
    journal = "Astron. Astrophys.",
    volume = "668",
    pages = "A188",
    year = "2022"
}

@ARTICLE{Bullock2017ARAA55343B,
       author = {{Bullock}, James S. and {Boylan-Kolchin}, Michael},
        title = "{Small-Scale Challenges to the {\ensuremath{\Lambda}}CDM Paradigm}",
      journal = {\araa},
         year = 2017,
        month = aug,
       volume = {55},
       number = {1},
        pages = {343-387},
          doi = {10.1146/annurev-astro-091916-055313},
archivePrefix = {arXiv},
       eprint = {1707.04256},
 primaryClass = {astro-ph.CO},
       adsurl = {https://ui.adsabs.harvard.edu/abs/2017ARA&A..55..343B}
}

@ARTICLE{Weinberg2015PNAS11212249W,
       author = {{Weinberg}, David H. and {Bullock}, James S. and {Governato}, Fabio and {Kuzio de Naray}, Rachel and {Peter}, Annika H.~G.},
        title = "{Cold dark matter: Controversies on small scales}",
      journal = {Proceedings of the National Academy of Science},
         year = 2015,
        month = oct,
       volume = {112},
       number = {40},
        pages = {12249-12255},
          doi = {10.1073/pnas.1308716112},
archivePrefix = {arXiv},
       eprint = {1306.0913},
 primaryClass = {astro-ph.CO},
       adsurl = {https://ui.adsabs.harvard.edu/abs/2015PNAS..11212249W}
}

@ARTICLE{ElZant2001ApJ560636E,
       author = {{El-Zant}, Amr and {Shlosman}, Isaac and {Hoffman}, Yehuda},
        title = "{Dark Halos: The Flattening of the Density Cusp by Dynamical Friction}",
      journal = {\apj},
         year = 2001,
        month = oct,
       volume = {560},
       number = {2},
        pages = {636-643},
          doi = {10.1086/322516},
archivePrefix = {arXiv},
       eprint = {astro-ph/0103386},
 primaryClass = {astro-ph},
       adsurl = {https://ui.adsabs.harvard.edu/abs/2001ApJ...560..636E}
}

@ARTICLE{Goerdt2010ApJ7251707G,
       author = {{Goerdt}, Tobias and {Moore}, Ben and {Read}, J.~I. and {Stadel}, Joachim},
        title = "{Core Creation in Galaxies and Halos Via Sinking Massive Objects}",
      journal = {\apj},
         year = 2010,
        month = dec,
       volume = {725},
       number = {2},
        pages = {1707-1716},
          doi = {10.1088/0004-637X/725/2/1707},
archivePrefix = {arXiv},
       eprint = {0806.1951},
 primaryClass = {astro-ph},
       adsurl = {https://ui.adsabs.harvard.edu/abs/2010ApJ...725.1707G}
}

@ARTICLE{Pontzen2012MNRAS4213464P,
       author = {{Pontzen}, Andrew and {Governato}, Fabio},
        title = "{How supernova feedback turns dark matter cusps into cores}",
      journal = {\mnras},
         year = 2012,
        month = apr,
       volume = {421},
       number = {4},
        pages = {3464-3471},
          doi = {10.1111/j.1365-2966.2012.20571.x},
archivePrefix = {arXiv},
       eprint = {1106.0499},
 primaryClass = {astro-ph.CO},
       adsurl = {https://ui.adsabs.harvard.edu/abs/2012MNRAS.421.3464P}
}

@ARTICLE{Teyssier2013MNRAS4293068T,
       author = {{Teyssier}, Romain and {Pontzen}, Andrew and {Dubois}, Yohan and {Read}, Justin I.},
        title = "{Cusp-core transformations in dwarf galaxies: observational predictions}",
      journal = {\mnras},
         year = 2013,
        month = mar,
       volume = {429},
       number = {4},
        pages = {3068-3078},
          doi = {10.1093/mnras/sts563},
archivePrefix = {arXiv},
       eprint = {1206.4895},
 primaryClass = {astro-ph.CO},
       adsurl = {https://ui.adsabs.harvard.edu/abs/2013MNRAS.429.3068T}
}

@ARTICLE{Freundlich2020MNRAS4914523F,
       author = {{Freundlich}, Jonathan and {Dekel}, Avishai and {Jiang}, Fangzhou and {Ishai}, Guy and {Cornuault}, Nicolas and {Lapiner}, Sharon and {Dutton}, Aaron A. and {Macci{\`o}}, Andrea V.},
        title = "{A model for core formation in dark matter haloes and ultra-diffuse galaxies by outflow episodes}",
      journal = {\mnras},
         year = 2020,
        month = jan,
       volume = {491},
       number = {3},
        pages = {4523-4542},
          doi = {10.1093/mnras/stz3306},
archivePrefix = {arXiv},
       eprint = {1907.11726},
 primaryClass = {astro-ph.GA},
       adsurl = {https://ui.adsabs.harvard.edu/abs/2020MNRAS.491.4523F}
}

@ARTICLE{Ostriker2019ApJ88597O,
       author = {{Ostriker}, Jeremiah P. and {Choi}, Ena and {Chow}, Anthony and {Guha}, Kundan},
        title = "{Mind the Gap: Is the Too Big to Fail Problem Resolved?}",
      journal = {\apj},
         year = 2019,
        month = nov,
       volume = {885},
       number = {1},
          eid = {97},
        pages = {97},
          doi = {10.3847/1538-4357/ab3288},
archivePrefix = {arXiv},
       eprint = {1904.10471},
 primaryClass = {astro-ph.GA},
       adsurl = {https://ui.adsabs.harvard.edu/abs/2019ApJ...885...97O}
}

@ARTICLE{Verbeke2017AA607A13V,
       author = {{Verbeke}, R. and {Papastergis}, E. and {Ponomareva}, A.~A. and {Rathi}, S. and {De Rijcke}, S.},
        title = "{A new astrophysical solution to the Too Big To Fail problem. Insights from the moria simulations}",
      journal = {\aap},
         year = 2017,
        month = oct,
       volume = {607},
          eid = {A13},
        pages = {A13},
          doi = {10.1051/0004-6361/201730758},
archivePrefix = {arXiv},
       eprint = {1703.03810},
 primaryClass = {astro-ph.GA},
       adsurl = {https://ui.adsabs.harvard.edu/abs/2017A&A...607A..13V}
}

@ARTICLE{Lovell2017MNRAS4682836L,
       author = {{Lovell}, Mark R. and {Gonzalez-Perez}, Violeta and {Bose}, Sownak and {Boyarsky}, Alexey and {Cole}, Shaun and {Frenk}, Carlos S. and {Ruchayskiy}, Oleg},
        title = "{Addressing the too big to fail problem with baryon physics and sterile neutrino dark matter}",
      journal = {\mnras},
         year = 2017,
        month = jul,
       volume = {468},
       number = {3},
        pages = {2836-2849},
          doi = {10.1093/mnras/stx621},
archivePrefix = {arXiv},
       eprint = {1611.00005},
 primaryClass = {astro-ph.GA},
       adsurl = {https://ui.adsabs.harvard.edu/abs/2017MNRAS.468.2836L}
}

@ARTICLE{Tomozeiu2016ApJ827L15T,
       author = {{Tomozeiu}, Mihai and {Mayer}, Lucio and {Quinn}, Thomas},
        title = "{Tidal Stirring of Satellites with Shallow Density Profiles Prevents Them from Being Too Big to Fail}",
      journal = {\apjl},
         year = 2016,
        month = aug,
       volume = {827},
       number = {1},
          eid = {L15},
        pages = {L15},
          doi = {10.3847/2041-8205/827/1/L15},
archivePrefix = {arXiv},
       eprint = {1605.00004},
 primaryClass = {astro-ph.GA},
       adsurl = {https://ui.adsabs.harvard.edu/abs/2016ApJ...827L..15T}
}

@ARTICLE{Homma2024PASJ76733H,
       author = {{Homma}, Daisuke and {Chiba}, Masashi and {Komiyama}, Yutaka and {Tanaka}, Masayuki and {Okamoto}, Sakurako and {Tanaka}, Mikito and {Ishigaki}, Miho N. and {Hayashi}, Kohei and {Arimoto}, Nobuo and {Lupton}, Robert H. and {Strauss}, Michael A. and {Miyazaki}, Satoshi and {Wang}, Shiang-Yu and {Murayama}, Hitoshi},
        title = "{Final results of the search for new Milky Way satellites in the Hyper Suprime-Cam Subaru Strategic Program survey: Discovery of two more candidates}",
      journal = {\pasj},
         year = 2024,
        month = aug,
       volume = {76},
       number = {4},
        pages = {733-752},
          doi = {10.1093/pasj/psae044},
archivePrefix = {arXiv},
       eprint = {2311.05439},
 primaryClass = {astro-ph.GA},
       adsurl = {https://ui.adsabs.harvard.edu/abs/2024PASJ...76..733H}
}

@ARTICLE{Brooks2013ApJ76522B,
       author = {{Brooks}, Alyson M. and {Kuhlen}, Michael and {Zolotov}, Adi and {Hooper}, Dan},
        title = "{A Baryonic Solution to the Missing Satellites Problem}",
      journal = {\apj},
         year = 2013,
        month = mar,
       volume = {765},
       number = {1},
          eid = {22},
        pages = {22},
          doi = {10.1088/0004-637X/765/1/22},
archivePrefix = {arXiv},
       eprint = {1209.5394},
 primaryClass = {astro-ph.CO},
       adsurl = {https://ui.adsabs.harvard.edu/abs/2013ApJ...765...22B}
}

@ARTICLE{Sales2022NatAs6897S,
       author = {{Sales}, Laura V. and {Wetzel}, Andrew and {Fattahi}, Azadeh},
        title = "{Baryonic solutions and challenges for cosmological models of dwarf galaxies}",
      journal = {Nature Astronomy},
         year = 2022,
        month = jun,
       volume = {6},
        pages = {897-910},
          doi = {10.1038/s41550-022-01689-w},
archivePrefix = {arXiv},
       eprint = {2206.05295},
 primaryClass = {astro-ph.GA},
       adsurl = {https://ui.adsabs.harvard.edu/abs/2022NatAs...6..897S}
}

@ARTICLE{Sawala2023NatAs7481S,
       author = {{Sawala}, Till and {Cautun}, Marius and {Frenk}, Carlos and {Helly}, John and {Jasche}, Jens and {Jenkins}, Adrian and {Johansson}, Peter H. and {Lavaux}, Guilhem and {McAlpine}, Stuart and {Schaller}, Matthieu},
        title = "{The Milky Way's plane of satellites is consistent with {\ensuremath{\Lambda}}CDM}",
      journal = {Nature Astronomy},
         year = 2023,
        month = apr,
       volume = {7},
        pages = {481-491},
          doi = {10.1038/s41550-022-01856-z},
archivePrefix = {arXiv},
       eprint = {2205.02860},
 primaryClass = {astro-ph.GA},
       adsurl = {https://ui.adsabs.harvard.edu/abs/2023NatAs...7..481S}
}

@ARTICLE{Sawala2016MNRAS4571931S,
       author = {{Sawala}, Till and {Frenk}, Carlos S. and {Fattahi}, Azadeh and {Navarro}, Julio F. and {Bower}, Richard G. and {Crain}, Robert A. and {Dalla Vecchia}, Claudio and {Furlong}, Michelle and {Helly}, John. C. and {Jenkins}, Adrian and {Oman}, Kyle A. and {Schaller}, Matthieu and {Schaye}, Joop and {Theuns}, Tom and {Trayford}, James and {White}, Simon D.~M.},
        title = "{The APOSTLE simulations: solutions to the Local Group's cosmic puzzles}",
      journal = {\mnras},
         year = 2016,
        month = apr,
       volume = {457},
       number = {2},
        pages = {1931-1943},
          doi = {10.1093/mnras/stw145},
archivePrefix = {arXiv},
       eprint = {1511.01098},
 primaryClass = {astro-ph.GA},
       adsurl = {https://ui.adsabs.harvard.edu/abs/2016MNRAS.457.1931S}
}

@article{Tokayer:2024wwo,
    author = "Tokayer, Yarone M. and Dutra, Isaque and Natarajan, Priyamvada and Mahler, Guillaume and Jauzac, Mathilde and Meneghetti, Massimo",
    title = "{The Galaxy\textendash{}Galaxy Strong Lensing Cross Section and the Internal Distribution of Matter in \ensuremath{\Lambda}CDM Substructure}",
    eprint = "2404.16951",
    archivePrefix = "arXiv",
    primaryClass = "astro-ph.CO",
    doi = "10.3847/1538-4357/ad51fd",
    journal = "Astrophys. J.",
    volume = "970",
    number = "2",
    pages = "143",
    year = "2024"
}

@article{Ragagnin:2022usu,
    author = "Ragagnin, Antonio and others",
    title = "{Galaxies in the central regions of simulated galaxy clusters}",
    eprint = "2204.09067",
    archivePrefix = "arXiv",
    primaryClass = "astro-ph.CO",
    doi = "10.1051/0004-6361/202243651",
    journal = "Astron. Astrophys.",
    volume = "665",
    pages = "A16",
    year = "2022"
}

@article{Aboubrahim:2020lnr,
    author = "Aboubrahim, Amin and Feng, Wan-Zhe and Nath, Pran and Wang, Zhu-Yao",
    title = "{Self-interacting hidden sector dark matter, small scale galaxy structure anomalies, and a dark force}",
    eprint = "2008.00529",
    archivePrefix = "arXiv",
    primaryClass = "hep-ph",
    doi = "10.1103/PhysRevD.103.075014",
    journal = "Phys. Rev. D",
    volume = "103",
    number = "7",
    pages = "075014",
    year = "2021"
}

@article{Feng:2009mn,
    author = "Feng, Jonathan L. and Kaplinghat, Manoj and Tu, Huitzu and Yu, Hai-Bo",
    title = "{Hidden Charged Dark Matter}",
    eprint = "0905.3039",
    archivePrefix = "arXiv",
    primaryClass = "hep-ph",
    reportNumber = "UCI-TR-2009-06",
    doi = "10.1088/1475-7516/2009/07/004",
    journal = "JCAP",
    volume = "07",
    pages = "004",
    year = "2009"
}

@article{Tulin:2012wi,
    author = "Tulin, Sean and Yu, Hai-Bo and Zurek, Kathryn M.",
    title = "{Resonant Dark Forces and Small Scale Structure}",
    eprint = "1210.0900",
    archivePrefix = "arXiv",
    primaryClass = "hep-ph",
    reportNumber = "MCTP-12-27",
    doi = "10.1103/PhysRevLett.110.111301",
    journal = "Phys. Rev. Lett.",
    volume = "110",
    number = "11",
    pages = "111301",
    year = "2013"
}

@ARTICLE{Eichner2013ApJ774124E,
       author = {{Eichner}, Thomas and {Seitz}, Stella and {Suyu}, Sherry H. and {Halkola}, Aleksi and {Umetsu}, Keiichi and {Zitrin}, Adi and {Coe}, Dan and {Monna}, Anna and {Rosati}, Piero and {Grillo}, Claudio and {Balestra}, Italo and {Postman}, Marc and {Koekemoer}, Anton and {Zheng}, Wei and {H{\o}st}, Ole and {Lemze}, Doron and {Broadhurst}, Tom and {Moustakas}, Leonidas and {Bradley}, Larry and {Molino}, Alberto and {Nonino}, Mario and {Mercurio}, Amata and {Scodeggio}, Marco and {Bartelmann}, Matthias and {Benitez}, Narciso and {Bouwens}, Rychard and {Donahue}, Megan and {Infante}, Leopoldo and {Jouvel}, Stephanie and {Kelson}, Daniel and {Lahav}, Ofer and {Medezinski}, Elinor and {Melchior}, Peter and {Merten}, Julian and {Riess}, Adam},
        title = "{Galaxy Halo Truncation and Giant Arc Surface Brightness Reconstruction in the Cluster MACSJ1206.2-0847}",
      journal = {\apj},
         year = 2013,
        month = sep,
       volume = {774},
       number = {2},
          eid = {124},
        pages = {124},
          doi = {10.1088/0004-637X/774/2/124},
archivePrefix = {arXiv},
       eprint = {1306.5240},
 primaryClass = {astro-ph.CO},
       adsurl = {https://ui.adsabs.harvard.edu/abs/2013ApJ...774..124E}
}

@ARTICLE{Umetsu2014ApJ795163U,
       author = {{Umetsu}, Keiichi and {Medezinski}, Elinor and {Nonino}, Mario and {Merten}, Julian and {Postman}, Marc and {Meneghetti}, Massimo and {Donahue}, Megan and {Czakon}, Nicole and {Molino}, Alberto and {Seitz}, Stella and {Gruen}, Daniel and {Lemze}, Doron and {Balestra}, Italo and {Ben{\'\i}tez}, Narciso and {Biviano}, Andrea and {Broadhurst}, Tom and {Ford}, Holland and {Grillo}, Claudio and {Koekemoer}, Anton and {Melchior}, Peter and {Mercurio}, Amata and {Moustakas}, John and {Rosati}, Piero and {Zitrin}, Adi},
        title = "{CLASH: Weak-lensing Shear-and-magnification Analysis of 20 Galaxy Clusters}",
      journal = {\apj},
         year = 2014,
        month = nov,
       volume = {795},
       number = {2},
          eid = {163},
        pages = {163},
          doi = {10.1088/0004-637X/795/2/163},
archivePrefix = {arXiv},
       eprint = {1404.1375},
 primaryClass = {astro-ph.CO},
       adsurl = {https://ui.adsabs.harvard.edu/abs/2014ApJ...795..163U}
}

@ARTICLE{Coe2019ApJ88485C,
       author = {{Coe}, Dan and {Salmon}, Brett and {Brada{\v{c}}}, Maru{\v{s}}a and {Bradley}, Larry D. and {Sharon}, Keren and {Zitrin}, Adi and {Acebron}, Ana and {Cerny}, Catherine and {Cibirka}, Nath{\'a}lia and {Strait}, Victoria and {Paterno-Mahler}, Rachel and {Mahler}, Guillaume and {Avila}, Roberto J. and {Ogaz}, Sara and {Huang}, Kuang-Han and {Pelliccia}, Debora and {Stark}, Daniel P. and {Mainali}, Ramesh and {Oesch}, Pascal A. and {Trenti}, Michele and {Carrasco}, Daniela and {Dawson}, William A. and {Rodney}, Steven A. and {Strolger}, Louis-Gregory and {Riess}, Adam G. and {Jones}, Christine and {Frye}, Brenda L. and {Czakon}, Nicole G. and {Umetsu}, Keiichi and {Vulcani}, Benedetta and {Graur}, Or and {Jha}, Saurabh W. and {Graham}, Melissa L. and {Molino}, Alberto and {Nonino}, Mario and {Hjorth}, Jens and {Selsing}, Jonatan and {Christensen}, Lise and {Kikuchihara}, Shotaro and {Ouchi}, Masami and {Oguri}, Masamune and {Welch}, Brian and {Lemaux}, Brian C. and {Andrade-Santos}, Felipe and {Hoag}, Austin T. and {Johnson}, Traci L. and {Peterson}, Avery and {Past}, Matthew and {Fox}, Carter and {Agulli}, Irene and {Livermore}, Rachael and {Ryan}, Russell E. and {Lam}, Daniel and {Sendra-Server}, Irene and {Toft}, Sune and {Lovisari}, Lorenzo and {Su}, Yuanyuan},
        title = "{RELICS: Reionization Lensing Cluster Survey}",
      journal = {\apj},
         year = 2019,
        month = oct,
       volume = {884},
       number = {1},
          eid = {85},
        pages = {85},
          doi = {10.3847/1538-4357/ab412b},
archivePrefix = {arXiv},
       eprint = {1903.02002},
 primaryClass = {astro-ph.GA},
       adsurl = {https://ui.adsabs.harvard.edu/abs/2019ApJ...884...85C}
}

@ARTICLE{Cerny2018ApJ859159C,
       author = {{Cerny}, Catherine and {Sharon}, Keren and {Andrade-Santos}, Felipe and {Avila}, Roberto J. and {Brada{\v{c}}}, Maru{\v{s}}a and {Bradley}, Larry D. and {Carrasco}, Daniela and {Coe}, Dan and {Czakon}, Nicole G. and {Dawson}, William A. and {Frye}, Brenda L. and {Hoag}, Austin and {Huang}, Kuang-Han and {Johnson}, Traci L. and {Jones}, Christine and {Lam}, Daniel and {Lovisari}, Lorenzo and {Mainali}, Ramesh and {Oesch}, Pascal A. and {Ogaz}, Sara and {Past}, Matthew and {Paterno-Mahler}, Rachel and {Peterson}, Avery and {Riess}, Adam G. and {Rodney}, Steven A. and {Ryan}, Russell E. and {Salmon}, Brett and {Sendra-Server}, Irene and {Stark}, Daniel P. and {Strolger}, Louis-Gregory and {Trenti}, Michele and {Umetsu}, Keiichi and {Vulcani}, Benedetta and {Zitrin}, Adi},
        title = "{RELICS: Strong Lens Models for Five Galaxy Clusters from the Reionization Lensing Cluster Survey}",
      journal = {\apj},
         year = 2018,
        month = jun,
       volume = {859},
       number = {2},
          eid = {159},
        pages = {159},
          doi = {10.3847/1538-4357/aabe7b},
archivePrefix = {arXiv},
       eprint = {1710.09329},
 primaryClass = {astro-ph.GA},
       adsurl = {https://ui.adsabs.harvard.edu/abs/2018ApJ...859..159C}
}

@ARTICLE{Natarajan2024SSRv22019N,
       author = {{Natarajan}, P. and {Williams}, L.~L.~R. and {Brada{\v{c}}}, M. and {Grillo}, C. and {Ghosh}, A. and {Sharon}, K. and {Wagner}, J.},
        title = "{Strong Lensing by Galaxy Clusters}",
      journal = {\ssr},
         year = 2024,
        month = feb,
       volume = {220},
       number = {2},
          eid = {19},
        pages = {19},
          doi = {10.1007/s11214-024-01051-8},
archivePrefix = {arXiv},
       eprint = {2403.06245},
 primaryClass = {astro-ph.CO},
       adsurl = {https://ui.adsabs.harvard.edu/abs/2024SSRv..220...19N}
}

@article{Moore:2000fp,
    author = "Moore, Ben and Gelato, Sergio and Jenkins, Adrian and Pearce, F. R. and Quilis, Vicent",
    title = "{Collisional versus collisionless dark matter}",
    eprint = "astro-ph/0002308",
    archivePrefix = "arXiv",
    doi = "10.1086/312692",
    journal = "Astrophys. J. Lett.",
    volume = "535",
    pages = "L21--L24",
    year = "2000"
}

@article{Nadler:2020ulu,
    author = "Nadler, Ethan O. and Banerjee, Arka and Adhikari, Susmita and Mao, Yao-Yuan and Wechsler, Risa H.",
    title = "{Signatures of Velocity-Dependent Dark Matter Self-Interactions in Milky Way-mass Halos}",
    eprint = "2001.08754",
    archivePrefix = "arXiv",
    primaryClass = "astro-ph.CO",
    doi = "10.3847/1538-4357/ab94b0",
    journal = "Astrophys. J.",
    volume = "896",
    number = "2",
    pages = "112",
    year = "2020"
}

@ARTICLE{Ghigna1998MNRAS300146G,
       author = {{Ghigna}, Sebastiano and {Moore}, Ben and {Governato}, Fabio and {Lake}, George and {Quinn}, Thomas and {Stadel}, Joachim},
        title = "{Dark matter haloes within clusters}",
      journal = {\mnras},
         year = 1998,
        month = oct,
       volume = {300},
       number = {1},
        pages = {146-162},
          doi = {10.1046/j.1365-8711.1998.01918.x},
archivePrefix = {arXiv},
       eprint = {astro-ph/9801192},
 primaryClass = {astro-ph},
       adsurl = {https://ui.adsabs.harvard.edu/abs/1998MNRAS.300..146G}
}

@ARTICLE{Taylor2001ApJ559716T,
       author = {{Taylor}, James E. and {Babul}, Arif},
        title = "{The Dynamics of Sinking Satellites around Disk Galaxies: A Poor Man's Alternative to High-Resolution Numerical Simulations}",
      journal = {\apj},
         year = 2001,
        month = oct,
       volume = {559},
       number = {2},
        pages = {716-735},
          doi = {10.1086/322276},
archivePrefix = {arXiv},
       eprint = {astro-ph/0012305},
 primaryClass = {astro-ph},
       adsurl = {https://ui.adsabs.harvard.edu/abs/2001ApJ...559..716T}
}

@ARTICLE{Wu2013ApJ76723W,
       author = {{Wu}, Hao-Yi and {Hahn}, Oliver and {Wechsler}, Risa H. and {Behroozi}, Peter S. and {Mao}, Yao-Yuan},
        title = "{Rhapsody. II. Subhalo Properties and the Impact of Tidal Stripping From a Statistical Sample of Cluster-size Halos}",
      journal = {\apj},
         year = 2013,
        month = apr,
       volume = {767},
       number = {1},
          eid = {23},
        pages = {23},
          doi = {10.1088/0004-637X/767/1/23},
archivePrefix = {arXiv},
       eprint = {1210.6358},
 primaryClass = {astro-ph.CO},
       adsurl = {https://ui.adsabs.harvard.edu/abs/2013ApJ...767...23W}
}

@ARTICLE{Hahn2009MNRAS3981742H,
       author = {{Hahn}, Oliver and {Porciani}, Cristiano and {Dekel}, Avishai and {Carollo}, C. Marcella},
        title = "{Tidal effects and the environment dependence of halo assembly}",
      journal = {\mnras},
         year = 2009,
        month = oct,
       volume = {398},
       number = {4},
        pages = {1742-1756},
          doi = {10.1111/j.1365-2966.2009.15271.x},
archivePrefix = {arXiv},
       eprint = {0803.4211},
 primaryClass = {astro-ph},
       adsurl = {https://ui.adsabs.harvard.edu/abs/2009MNRAS.398.1742H}
}

@ARTICLE{Green2021MNRAS5034075G,
       author = {{Green}, Sheridan B. and {van den Bosch}, Frank C. and {Jiang}, Fangzhou},
        title = "{The tidal evolution of dark matter substructure - II. The impact of artificial disruption on subhalo mass functions and radial profiles}",
      journal = {\mnras},
         year = 2021,
        month = may,
       volume = {503},
       number = {3},
        pages = {4075-4091},
          doi = {10.1093/mnras/stab696},
archivePrefix = {arXiv},
       eprint = {2103.01227},
 primaryClass = {astro-ph.GA},
       adsurl = {https://ui.adsabs.harvard.edu/abs/2021MNRAS.503.4075G}
}

@ARTICLE{Penarrubia2010MNRAS4061290P,
       author = {{Pe{\~n}arrubia}, Jorge and {Benson}, Andrew J. and {Walker}, Matthew G. and {Gilmore}, Gerard and {McConnachie}, Alan W. and {Mayer}, Lucio},
        title = "{The impact of dark matter cusps and cores on the satellite galaxy population around spiral galaxies}",
      journal = {\mnras},
         year = 2010,
        month = aug,
       volume = {406},
       number = {2},
        pages = {1290-1305},
          doi = {10.1111/j.1365-2966.2010.16762.x},
archivePrefix = {arXiv},
       eprint = {1002.3376},
 primaryClass = {astro-ph.GA},
       adsurl = {https://ui.adsabs.harvard.edu/abs/2010MNRAS.406.1290P}
}

@ARTICLE{vandenBosch2018MNRAS4754066V,
       author = {{van den Bosch}, Frank C. and {Ogiya}, Go},
        title = "{Dark matter substructure in numerical simulations: a tale of discreteness noise, runaway instabilities, and artificial disruption}",
      journal = {\mnras},
         year = 2018,
        month = apr,
       volume = {475},
       number = {3},
        pages = {4066-4087},
          doi = {10.1093/mnras/sty084},
archivePrefix = {arXiv},
       eprint = {1801.05427},
 primaryClass = {astro-ph.GA},
       adsurl = {https://ui.adsabs.harvard.edu/abs/2018MNRAS.475.4066V}
}

@ARTICLE{King.62,
       author = {{King}, Ivan},
        title = "{The structure of star clusters. I. an empirical density law}",
      journal = {\aj},
         year = 1962,
        month = oct,
       volume = {67},
        pages = {471},
          doi = {10.1086/108756},
       adsurl = {https://ui.adsabs.harvard.edu/abs/1962AJ.....67..471K}
}

@ARTICLE{Tollet.etal.17,
       author = {{Tollet}, {\'E}douard and {Cattaneo}, Andrea and {Mamon}, Gary A. and {Moutard}, Thibaud and {van den Bosch}, Frank C.},
        title = "{On stellar mass loss from galaxies in groups and clusters}",
      journal = {\mnras},
         year = 2017,
        month = nov,
       volume = {471},
       number = {4},
        pages = {4170-4193},
          doi = {10.1093/mnras/stx1840},
archivePrefix = {arXiv},
       eprint = {1707.06264},
 primaryClass = {astro-ph.GA},
       adsurl = {https://ui.adsabs.harvard.edu/abs/2017MNRAS.471.4170T}
}

@ARTICLE{vandenBosch2018MNRAS4743043V,
       author = {{van den Bosch}, Frank C. and {Ogiya}, Go and {Hahn}, Oliver and {Burkert}, Andreas},
        title = "{Disruption of dark matter substructure: fact or fiction?}",
      journal = {\mnras},
         year = 2018,
        month = mar,
       volume = {474},
       number = {3},
        pages = {3043-3066},
          doi = {10.1093/mnras/stx2956},
archivePrefix = {arXiv},
       eprint = {1711.05276},
 primaryClass = {astro-ph.GA},
       adsurl = {https://ui.adsabs.harvard.edu/abs/2018MNRAS.474.3043V}
}

@ARTICLE{Benson2022MNRAS5171398B,
       author = {{Benson}, Andrew J. and {Du}, Xiaolong},
        title = "{Tidal tracks and artificial disruption of cold dark matter haloes}",
      journal = {\mnras},
         year = 2022,
        month = nov,
       volume = {517},
       number = {1},
        pages = {1398-1406},
          doi = {10.1093/mnras/stac2750},
archivePrefix = {arXiv},
       eprint = {2206.01842},
 primaryClass = {astro-ph.GA},
       adsurl = {https://ui.adsabs.harvard.edu/abs/2022MNRAS.517.1398B}
}

@ARTICLE{Vollmer2001ApJ561708V,
       author = {{Vollmer}, B. and {Cayatte}, V. and {Balkowski}, C. and {Duschl}, W.~J.},
        title = "{Ram Pressure Stripping and Galaxy Orbits: The Case of the Virgo Cluster}",
      journal = {\apj},
         year = 2001,
        month = nov,
       volume = {561},
       number = {2},
        pages = {708-726},
          doi = {10.1086/323368},
archivePrefix = {arXiv},
       eprint = {astro-ph/0107237},
 primaryClass = {astro-ph},
       adsurl = {https://ui.adsabs.harvard.edu/abs/2001ApJ...561..708V}
}

@ARTICLE{Mace2024arXiv240201604M,
       author = {{Mace}, Charlie and {Carton Zeng}, Zhichao and {Peter}, Annika H.~G. and {Du}, Xiaolong and {Yang}, Shengqi and {Benson}, Andrew and {Vogelsberger}, Mark},
        title = "{Convergence Tests of Self-Interacting Dark Matter Simulations}",
      journal = {arXiv e-prints},
         year = 2024,
        month = feb,
          eid = {arXiv:2402.01604},
        pages = {arXiv:2402.01604},
          doi = {10.48550/arXiv.2402.01604},
archivePrefix = {arXiv},
       eprint = {2402.01604},
 primaryClass = {astro-ph.GA},
       adsurl = {https://ui.adsabs.harvard.edu/abs/2024arXiv240201604M}
}

@ARTICLE{Springel2010MNRAS401791S,
       author = {{Springel}, Volker},
        title = "{E pur si muove: Galilean-invariant cosmological hydrodynamical simulations on a moving mesh}",
      journal = {\mnras},
         year = 2010,
        month = jan,
       volume = {401},
       number = {2},
        pages = {791-851},
          doi = {10.1111/j.1365-2966.2009.15715.x},
archivePrefix = {arXiv},
       eprint = {0901.4107},
 primaryClass = {astro-ph.CO},
       adsurl = {https://ui.adsabs.harvard.edu/abs/2010MNRAS.401..791S}
}

@ARTICLE{Mitra2024MNRAS5333647M,
       author = {{Mitra}, Kaustav and {van den Bosch}, Frank C. and {Lange}, Johannes U.},
        title = "{BASILISK II. Improved constraints on the galaxy-halo connection from satellite kinematics in SDSS}",
      journal = {\mnras},
         year = 2024,
        month = sep,
       volume = {533},
       number = {3},
        pages = {3647-3675},
          doi = {10.1093/mnras/stae2030},
archivePrefix = {arXiv},
       eprint = {2409.03105},
 primaryClass = {astro-ph.CO},
       adsurl = {https://ui.adsabs.harvard.edu/abs/2024MNRAS.533.3647M}
}

@ARTICLE{vandenBosch2019MNRAS4884984V,
       author = {{van den Bosch}, Frank C. and {Lange}, Johannes U. and {Zentner}, Andrew R.},
        title = "{Basilisk: Bayesian hierarchical inference of the galaxy-halo connection using satellite kinematics - I. Method and validation}",
      journal = {\mnras},
         year = 2019,
        month = oct,
       volume = {488},
       number = {4},
        pages = {4984-5013},
          doi = {10.1093/mnras/stz2017},
archivePrefix = {arXiv},
       eprint = {1908.07547},
 primaryClass = {astro-ph.CO},
       adsurl = {https://ui.adsabs.harvard.edu/abs/2019MNRAS.488.4984V}
}

@ARTICLE{Wojtak2013MNRAS4282407W,
       author = {{Wojtak}, Rados{\l}aw and {Mamon}, Gary A.},
        title = "{Physical properties underlying observed kinematics of satellite galaxies}",
      journal = {\mnras},
         year = 2013,
        month = jan,
       volume = {428},
       number = {3},
        pages = {2407-2417},
          doi = {10.1093/mnras/sts203},
archivePrefix = {arXiv},
       eprint = {1207.1647},
 primaryClass = {astro-ph.CO},
       adsurl = {https://ui.adsabs.harvard.edu/abs/2013MNRAS.428.2407W}
}

@ARTICLE{Benatov2006MNRAS370427B,
       author = {{Benatov}, L. and {Rines}, K. and {Natarajan}, P. and {Kravtsov}, A. and {Nagai}, D.},
        title = "{Galaxy orbits and the intracluster gas temperature in clusters}",
      journal = {\mnras},
         year = 2006,
        month = jul,
       volume = {370},
       number = {1},
        pages = {427-434},
          doi = {10.1111/j.1365-2966.2006.10490.x},
archivePrefix = {arXiv},
       eprint = {astro-ph/0605105},
 primaryClass = {astro-ph},
       adsurl = {https://ui.adsabs.harvard.edu/abs/2006MNRAS.370..427B}
}

@ARTICLE{OlaveRojas2018MNRAS4792328O,
       author = {{Olave-Rojas}, D. and {Cerulo}, P. and {Demarco}, R. and {Jaff{\'e}}, Y.~L. and {Mercurio}, A. and {Rosati}, P. and {Balestra}, I. and {Nonino}, M.},
        title = "{Galaxy pre-processing in substructures around z {\ensuremath{\sim}} 0.4 galaxy clusters}",
      journal = {\mnras},
         year = 2018,
        month = sep,
       volume = {479},
       number = {2},
        pages = {2328-2350},
          doi = {10.1093/mnras/sty1669},
archivePrefix = {arXiv},
       eprint = {1806.08435},
 primaryClass = {astro-ph.GA},
       adsurl = {https://ui.adsabs.harvard.edu/abs/2018MNRAS.479.2328O}
}

@ARTICLE{Muratov2010ApJ7181266M,
       author = {{Muratov}, Alexander L. and {Gnedin}, Oleg Y.},
        title = "{Modeling the Metallicity Distribution of Globular Clusters}",
      journal = {\apj},
         year = 2010,
        month = aug,
       volume = {718},
       number = {2},
        pages = {1266-1288},
          doi = {10.1088/0004-637X/718/2/1266},
archivePrefix = {arXiv},
       eprint = {1002.1325},
 primaryClass = {astro-ph.GA},
       adsurl = {https://ui.adsabs.harvard.edu/abs/2010ApJ...718.1266M}
}

@ARTICLE{Schwarz1978AnSta6461S,
       author = {{Schwarz}, Gideon},
        title = "{Estimating the Dimension of a Model}",
      journal = {Annals of Statistics},
         year = 1978,
        month = jul,
       volume = {6},
       number = {2},
        pages = {461-464},
       adsurl = {https://ui.adsabs.harvard.edu/abs/1978AnSta...6..461S}
}

@ARTICLE{Bhattacharyya2022ApJ93230B,
       author = {{Bhattacharyya}, Joy and {Adhikari}, Susmita and {Banerjee}, Arka and {More}, Surhud and {Kumar}, Amit and {Nadler}, Ethan O. and {Chatterjee}, Suchetana},
        title = "{The Signatures of Self-interacting Dark Matter and Subhalo Disruption on Cluster Substructure}",
      journal = {\apj},
         year = 2022,
        month = jun,
       volume = {932},
       number = {1},
          eid = {30},
        pages = {30},
          doi = {10.3847/1538-4357/ac68e9},
archivePrefix = {arXiv},
       eprint = {2106.08292},
 primaryClass = {astro-ph.CO},
       adsurl = {https://ui.adsabs.harvard.edu/abs/2022ApJ...932...30B}
}

@ARTICLE{Nelson2024AA686A157N,
       author = {{Nelson}, Dylan and {Pillepich}, Annalisa and {Ayromlou}, Mohammadreza and {Lee}, Wonki and {Lehle}, Katrin and {Rohr}, Eric and {Truong}, Nhut},
        title = "{Introducing the TNG-Cluster simulation: Overview and the physical properties of the gaseous intracluster medium}",
      journal = {\aap},
         year = 2024,
        month = jun,
       volume = {686},
          eid = {A157},
        pages = {A157},
          doi = {10.1051/0004-6361/202348608},
archivePrefix = {arXiv},
       eprint = {2311.06338},
 primaryClass = {astro-ph.GA},
       adsurl = {https://ui.adsabs.harvard.edu/abs/2024A&A...686A.157N}
}

@article{Buckley:2009in,
    author = "Buckley, Matthew R. and Fox, Patrick J.",
    title = "{Dark Matter Self-Interactions and Light Force Carriers}",
    eprint = "0911.3898",
    archivePrefix = "arXiv",
    primaryClass = "hep-ph",
    reportNumber = "FERMILAB-PUB-09-560-T",
    doi = "10.1103/PhysRevD.81.083522",
    journal = "Phys. Rev. D",
    volume = "81",
    pages = "083522",
    year = "2010"
}

@article{Tsai:2020vpi,
    author = "Tsai, Yu-Dai and McGehee, Robert and Murayama, Hitoshi",
    title = "{Resonant Self-Interacting Dark Matter from Dark QCD}",
    eprint = "2008.08608",
    archivePrefix = "arXiv",
    primaryClass = "hep-ph",
    reportNumber = "FERMILAB-PUB-20-365-AE-T",
    doi = "10.1103/PhysRevLett.128.172001",
    journal = "Phys. Rev. Lett.",
    volume = "128",
    number = "17",
    pages = "172001",
    year = "2022"
}

@article{Loeb:2010gj,
    author = "Loeb, Abraham and Weiner, Neal",
    title = "{Cores in Dwarf Galaxies from Dark Matter with a Yukawa Potential}",
    eprint = "1011.6374",
    archivePrefix = "arXiv",
    primaryClass = "astro-ph.CO",
    doi = "10.1103/PhysRevLett.106.171302",
    journal = "Phys. Rev. Lett.",
    volume = "106",
    pages = "171302",
    year = "2011"
}

@article{Chu:2018faw,
    author = "Chu, Xiaoyong and Garcia-Cely, Camilo and Murayama, Hitoshi",
    title = "{Finite-size dark matter and its effect on small-scale structure}",
    eprint = "1901.00075",
    archivePrefix = "arXiv",
    primaryClass = "hep-ph",
    reportNumber = "DESY-18-225, IPMU18-0207",
    doi = "10.1103/PhysRevLett.124.041101",
    journal = "Phys. Rev. Lett.",
    volume = "124",
    number = "4",
    pages = "041101",
    year = "2020"
}

@article{Chu:2018fzy,
    author = "Chu, Xiaoyong and Garcia-Cely, Camilo and Murayama, Hitoshi",
    title = "{Velocity Dependence from Resonant Self-Interacting Dark Matter}",
    eprint = "1810.04709",
    archivePrefix = "arXiv",
    primaryClass = "hep-ph",
    reportNumber = "DESY-18-176",
    doi = "10.1103/PhysRevLett.122.071103",
    journal = "Phys. Rev. Lett.",
    volume = "122",
    number = "7",
    pages = "071103",
    year = "2019"
}

\label{lastpage}
\end{document}